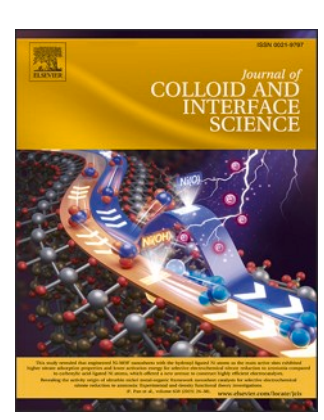

# Influence of chemistry and topography on the wettability of copper

Sarah Marie Lößlein [a,*], Rolf Merz [b], Yerila Rodríguez-Martínez [c,d], Florian Schäfer [e], Philipp G. Grützmacher [f], David Horwat [d], Michael Kopnarski [b], Frank Mücklich [a]

[a] Chair of Functional Materials, Department of Materials Science, Saarland University, 66123 Saarbrücken, Germany
[b] Institute for Surface and Thin Film Technologies (IFOS) at the University of Kaiserslautern-Landau (RPTU), Germany
[c] University of Havana, Photovoltaic Research Laboratory, Institute of Materials Science and Technology – Physics Faculty, San Lázaro y L, 10 400 Havana, Cuba
[d] Université de Lorraine, CNRS, IJL, F-54000 Nancy, France
[e] Materials Science and Methods, Department of Materials Science, Saarland University, 66123 Saarbrücken, Germany
[f] Institute for Engineering Design and Product Development, Tribology Research Division, TU Wien, 1060 Vienna, Austria

## GRAPHICAL ABSTRACT

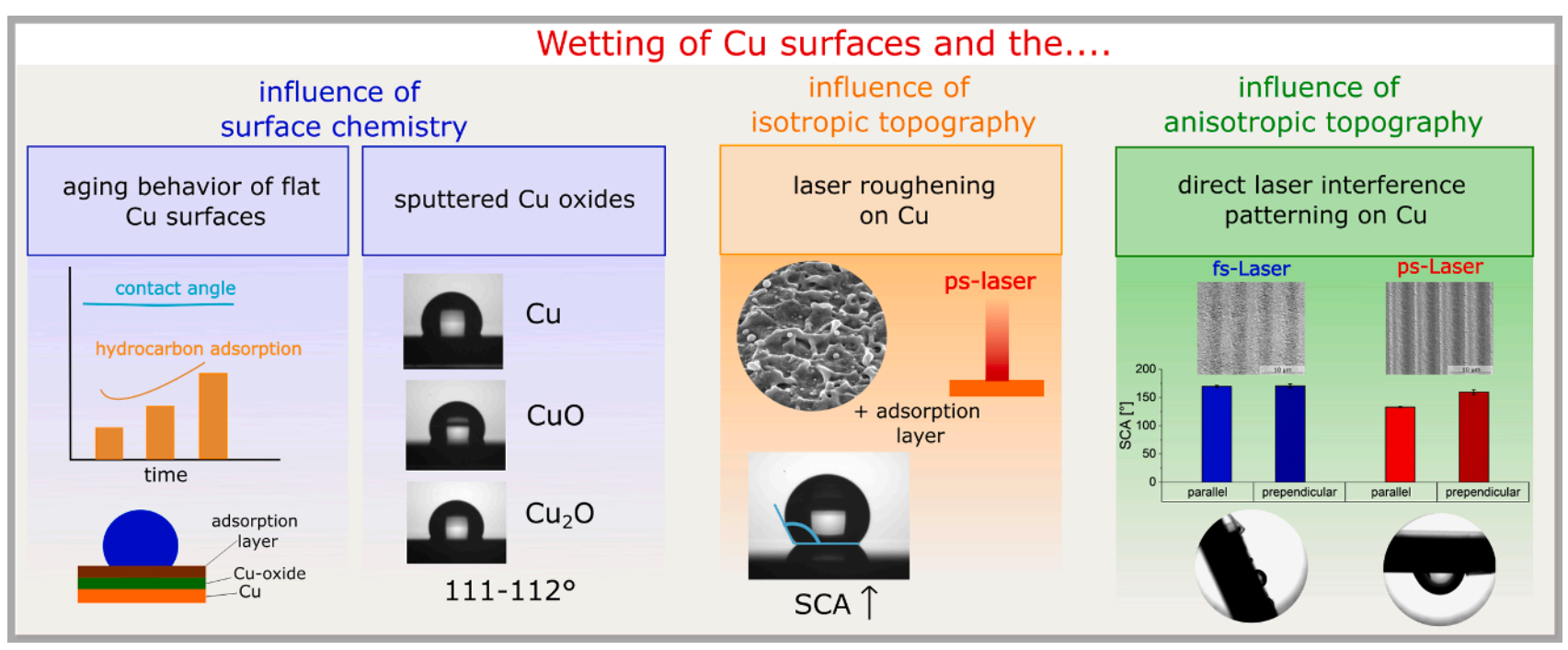

## ABSTRACT

To understand the complex interplay of topography and surface chemistry in wetting, fundamental studies investigating both parameters are needed. Due to the sensitivity of wetting to miniscule changes in one of the parameters it is imperative to precisely control the experimental approach. A profound understanding of their influence on wetting facilitates a tailored design of surfaces with unique functionality.

We present a multi-step study: The influence of surface chemistry is analyzed by determining the adsorption of volatile carbonous species (A) and by sputter deposition of metallic copper and copper oxides on flat copper substrates (B). A precise surface topography is created by laser processing. Isotropic topography is created by ps laser processing (C), and hierarchical anisotropic line patterns are produced by direct laser interference patterning (DLIP) with different pulse durations (D).

Our results reveal that the long-term wetting response of polished copper surfaces stabilizes with time despite ongoing accumulation of hydrocarbons and is dominated by this adsorption layer over the oxide state of the substrate (Cu, CuO, $Cu_2O$). The surfaces' wetting response can be precisely tuned by tailoring the topography via laser processing. The sub-pattern morphology of primary line-like patterns showed great impact on the static contact angle, wetting anisotropy, and water adhesion. An increased roughness inside the pattern valleys combined with a minor roughness on pattern peaks favors air-inclusions, isotropic hydrophobicity, and low water adhesion. Increasing depth of the primary topography can also induce air-inclusions despite increasing peak roughness while time dependent wetting transitions were observed.

* Corresponding author.
*E-mail address:* sarah.loesslein@uni-saarland.de (S. Marie Lößlein).

## 1. Introduction

The production of copper has been increasing continuously within recent years. The U.S. Geological Survey estimates an increase of approximately 30 % in the world wide copper production (by mining, smelter, and refinery) from 2010 to 2020 [1,2]. The Copper Development Association Inc. states that "Copper Is … The Metal of Civilization" [3]. Many of the unique properties of copper and its alloys that make copper so vital for a wide range of applications are linked to their wetting behavior. For instance, the antimicrobial efficiency depends on the wetting properties [4] as bacterial adhesion is influenced by the surface hydrophobicity [5,6]. Furthermore, Liu et al. [7] found that superhydrophobic coatings can improve the corrosion resistance of copper, which is especially important due to its frequent application in electrical connections. Wen et al. [8] produced hydrophobic copper nanowires to accelerate droplet removal for improvement of condensation heat transfer. Chemical treatments to grow copper oxides in combination with chemical hydrophobization also showed to increase heat transfer coefficients as well as critical heat flux [9]. Chemical surface functionalization of copper combined with laser treatment allows for the fabrication of superhydrophobic/hydrophilic hybrid surfaces to alter droplet condensation [10]. Kim et al. [11] equalized surface chemistry of copper samples by a hydrophobic coating to evaluate the influence of roughness on heat transfer. Unidirectional scratches were applied leading to wetting anisotropy but sandpaper scratching can induce very irregular topographies in contrast to laser patterning of surfaces and therefore limit the possibilities of roughness evaluation on wettability. Zupančič et al. [12] found indeed that laser patterning of steel foils and the formation of non-uniform patterns induces non-uniform wetting behavior which in turn greatly affects boiling heat transfer. Li et al. [13] altered copper chemically and topographically to increase hydrophobicity and improve self-cleaning behavior as well as water–oil separation performance.

The classic Young equation considers the contact angle on ideally smooth surfaces $\Theta_{ideal}$ following equation (1) [14] as a function of the surface tensions between liquid (L) and gas (G) $\sigma_{LG}$, solid (S) and gas $\sigma_{SG}$ as well as solid and liquid $\sigma_{SL}$:

$$\cos(\Theta_{ideal})^{*}\sigma_{LG} = \sigma_{SG} - \sigma_{SL} \tag{1}$$

Later approaches also considered the role of topography in the wetting behavior. Wenzel [15] postulated equation (2) for a system where the roughness is penetrated by the liquid. *r* represents the ratio of the area of the actual surface by the projected surface area and is $> 1$ for rough surfaces:

$$\cos(\Theta_{Wenzel}) = r^{*}\cos(\Theta_{ideal}) \tag{2}$$

Only for ideally smooth surfaces $r = 1$ and $\Theta_{Wenzel}$ transforms into $\Theta_{ideal}$ described in equation (1). Cassie and Baxter (CB) derived an equation for calculating the contact angle on rough surfaces for the case of water trapping air with the surface roughness [16]. While nowadays different forms of this equation find application in the scientific community, Milne et al. [17] recommend sticking to the original CB equation (3):

$$\cos(\Theta_{CB}) = f_1\cos(\Theta_{1ideal}) - f_2; \tag{3}$$

$f_1$ is defined as the area of the solid in contact with the liquid per projected area of the unit cell, $\Theta_{1ideal}$ as the contact angle on the smooth reference surface of material 1, while the subscript 2 represents area fraction of air (with an analogous contact angle of 180°).

In recent years, besides topography, the surface chemistry and its effect on wetting arose growing interest in the investigation of the wetting behavior of surfaces under ambient conditions. Different studies showed that the adsorption of volatile carbonous species leads to a transition of a hydrophilic behavior right after preparation of metallic and/or oxidic surfaces (e.g., by laser processing or simple grinding and polishing) to an increasingly hydrophobic behavior over time [18–21]. The earliest representation of a technical surface in contact with the atmosphere we were able to find is depicted in a review from J. E. Castle [22] and originally stems from a symposium contribution by F.R. Eirich [23] in 1968 (primary source was not accessible). He depicted a layered system of the metal surface with a growing oxide layer, an $OH^{-}$ termination of the oxide induced by atmospheric $H_2O$ and a subsequent attachment of polar and finally non-polar organic groups rendering the surface hydrophobic. Long et al. [18] did pioneering work in this field showing that storage of laser processed copper surfaces in organic rich atmospheres significantly increases the contact angle. More recent studies also showed that not only the atmosphere, but even sample packaging can lead to great deviations in the resulting contact angles [24,25]. Korczeniewski et al. [21] conducted a systematic study of the wetting transition of copper, based on controlled desorption and subsequent adsorption of hydrocarbons by exposing the surfaces to the atmosphere throughout 2 weeks and conducted simulations to explain the observed transition which showed that once a monolayer of hydrocarbon is formed, additional adsorption does not change the static contact angle significantly. In one of our previous works we were not able to find a saturation of adsorption within 30 days of storage [20]. In our current study, we present a long-term investigation of the adsorption of volatile carbon species throughout 43 weeks under controlled storage conditions and show a comparison to storage under ambient conditions.

Regarding the influence of the oxidation state of copper on the wetting behavior, recently Zhao et al. [26] showed that both CuO and $Cu_2O$ turn out hydrophobic upon exposure to C-rich atmosphere. The production of the two oxide states also introduced roughness to their surface and the storage time differed for both oxide types. Therefore, in our study we aimed at investigating the influence of oxide states (metallic Cu, CuO and $Cu_2O$) on flat samples stored under controlled conditions.

This approach allows for an assessment of the influence of surface chemistry separately from topography, which both have proven to influence the wetting behavior [27]. This is especially important for laser-processed copper surfaces, as both chemistry and topography form a complex interplay regarding the observed time-dependent wetting response. While different studies in literature investigated the wetting of laser-patterned copper surfaces [28–31], sample storage and packaging often are not or only vaguely ("ambient conditions") described. Also, in one of our previous studies [27], we stored laser patterned steel samples with and without PVD coating in ambient air and cleaned the samples in ethanol with the intention to create comparable surface conditions before measuring their wettability. We observed that both surface topography and chemistry are crucial for the wetting behavior. However, we also concluded that more detailed studies on both effects taking the adsorption of volatile carbon species into account will be necessary in the future to fundamentally understand the dependence of wetting properties of surfaces on these effects. While important insights have been gained during recent years into the interdependence of chemistry and topography on the wetting properties of copper surfaces, the detailed contributions are not clear yet. Therefore, in this work, we want to give a comprehensive overview of the wetting behavior of copper considering chemical as well as topographical surface modifications in a detailed multi-step analysis. Particularly, our approach takes the role of surface chemistry and of isotropic as well as anisotropic surface topography into consideration (compare Fig. 1). First, the aging behavior of bulk copper samples and the adsorption film forming was investigated. In a next step, the influence of the oxide state of copper on flat surfaces was investigated and finally to understand the influence of topography in the wetting behavior of copper, laser treatment was applied. First, isotropic roughness was created by ps laser treatment. Complex hierarchical patterns were created by direct laser interference patterning (DLIP) in a last step and their wetting response to water was analyzed based on the findings from the previous experiments.

## 2. Methods

### 2.1. Experimental Work

An oxygen-free copper sheet (CW008A, *Wieland*) of 1 mm thickness was mirror polished on an automated *Struers Tegra-Pol 21* with very light forces of 10 N in the final polishing steps of 3 μm and 1 μm as suggested in ref. [32] in order to achieve a mirror polished finish with an arithmetical mean height $S_a$ of 3–4 nm. Cleaning was performed by immersion in an ethanol ultrasonic bath for three minutes. Samples were dried by an ambient air stream.

### 2.2. Chemical surface treatment

Copper and copper oxides depositions were conducted by reactive DC magnetron sputtering. Polished copper samples were rinsed with ethanol, dried, and placed on a circular holder facing a copper target (purity of 99.99 %, Kurt J. Lesker Company) at a distance of 5 cm. The distance between the samples and the center of rotation of the holder (rotation frequency of 3.5 Hz) was 7.5 cm. After evacuation, argon was introduced with a flow rate of 50 sccm ($3.5*10^{-3}$ mbar of pressure) and sample surface etching was performed during 1 min by radiofrequency polarization. Parameters for the deposition of Cu and the oxides can be found in the supplementary material (Table S1).

### 2.3. Laser surface treatment

Isotropic laser roughening (LR) of the polished surfaces was conducted with an *Edgewave* (*PX50-1-GF)* PX-series laser in a single beam setup with a pulse duration of 12 ps and a wavelength of 532 nm. Additionally, direct laser interference patterning (DLIP) with two different lasers was employed to create line-like patterns. A femtosecond laser (*SOL ACE100F1K HP from Newport Spectra-Physics GmbH*) with a wavelength of 800 nm and a pulse duration of 100 fs was split into two beams by a diffractive optic element (DOE) and brought to interference subsequently on the sample surface in order to produce line patterns with a periodicity of 6 μm. The *Edgewave* picosecond laser was used equipped with an optical head from *Fraunhofer IWS Dresden* splitting the beam with a DOE, parallelizing the sub-beams with a prism and focusing on the sample surface to cause interference (see ref. [33] for a schematic overview of the beam path). A mask was installed to block two of the four sub-beams to create line-like patterns. Parameters for the laser patterning in the LR as well as the DLIP set up are summarized in Table S2 in the supplementary material. Details on sample storage for the individual sample sets can be found in the supplementary material (Table S3).

### 2.4. Surface characterization

For surface characterization, confocal laser scanning microscopy (LSM) was conducted on an *Olympus Lext OLS4100*. Analyzed roughness parameters include the arithmetical mean height $S_a$, the mean height of the profile elements $R_c$, the root-mean-square deviation $R_q$ and the S-ratio, which is defined as the ratio of the rough surface by the projected area and therewith resembles *r* in equation (2). Atomic force microscopy (AFM) was conducted on a *Bruker Dimension Icon*. Triangular-shaped silicon nitride SCANASYST AIR probes with a nominal radius of 2 nm were used for peak force tapping in the ScanAsyst Mode. For scans with a size of 5 x 5 $\mu m^2$ a scan rate of 0.248 Hz was applied and increased to 0.988 Hz for bigger scans of 20 x 20 $\mu m^2$. Imaging of the surface topography was performed by scanning electron microscopy (SEM) on a *FEI Helios FineLab600 FIB* at an acceleration voltage of 5 kV and 1.4nA beam current in the secondary electron mode. High resolution micrographs were taken with an in-lens detector at 5 kV and 86 pA. For focused ion-beam (FIB) cross-sectional characterization, platinum layers were sputter-deposited to avoid damage of the surface, like rounding of the edges during cutting with a 30 kV gallium ion beam. Chemical characterization was conducted by grazing incidence X-ray diffraction (XRD; PANalytcal X'Pert Pro-MPD; Cu $K_\alpha$ X-ray source; wavelength: 1.54 Å, 1° grazing angle). Samples were also characterized using a Bruker D8 Advanced diffractometer (Cu $K_\alpha$ X-ray source, wavelength: 1.54 Å) with a Bragg/Brentano configuration.

Surface chemistry was analyzed by X-Ray photoelectron spectroscopy (XPS). For this, an Axis Nova small spot electron spectrometer (Kratos Ltd.) was used. Three oblong spots of size 350 x 700 $\mu m^2$ per sample were chosen. All measurements were done with the detector axis set parallel to the surface normal of the sample. Only for a XPS tilt experiment, the detector axis was inclined from the 90° angle, measured between detector axis and surface, to 30° to reduce the information depth. By directly comparing the results of the two detector viewing directions, information concerning the elemental depth distribution can be concluded. Elemental concentrations were calculated from survey spectra (electron pass energy: 160 eV), using standard sensitivity factors, given by Kratos. Ltd. and Shirley shape fits for background subtraction. The binding states of detected elements are characterized by detailed core level spectra (electron pass energy: 20 eV) of the relevant elements (C1s, O1s, Cu2p). For Cu an additional detail spectrum of the Cu LMM Auger electron line was recorded. For details on the fitting procedure please refer to supplementary material. Additionally, XPS depth profiles were recorded, by sputter removal of surface layers with a 2 keV argon ion beam. The depth scale was calibrated according a $Ta_2O_5$ reference sample, giving a sputter velocity of around 3 nm/min.

For wetting analysis, static contact angles (SCA) were determined on a *Krüss DSA 100* equipped with a gastight Hamilton syringe (100 μl). On flat and LR samples, 3 μl droplets of purified water (spec. conductivity $\leq$ 1 μS/cm) were applied and images were taken 10 s after droplet deposition from two sides. On DLIP samples, droplet volume was increased to 6 μl to allow droplet deposition on the highly hydrophobic surfaces. Additionally, measurements were conducted approximately 60 s as well as 180 s after droplet deposition, where evaporation was suppressed by the cover set up introduced in an earlier work [24]. SCA results are portrayed as the mean of multiple measurements with error bars indicating the standard deviation. Atmospheric conditions were controlled during SCA measurements where details can be found in the respective results section (RH refers to relative humidity during droplet deposition). For top view images of droplets an extra set of droplets was applied by a pipette under a *VHX-7000 Keyence* optical microscope. In order to distinguish between a strong and a low water adhesion, a self-designed prototype tilting stage based on a step motor was used. DLIP samples were tilted with approximately 2°/s after application of 30 µl of water.

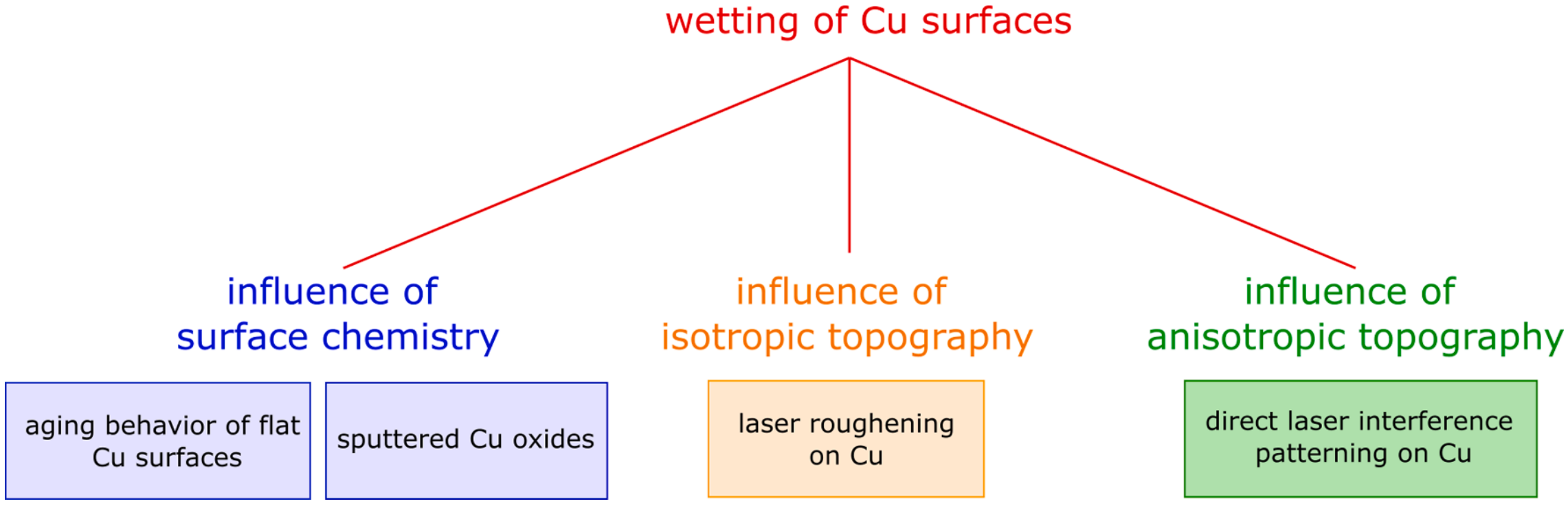

**Fig. 1.** Schematic describing the structure of the manuscript.

## 3. Results and discussion

### A) Aging of Copper samples

In a previous study we were able to show the strong influence of storage conditions on the development of the wetting behavior of mechanically polished copper surfaces [24]. The transition from hydrophilic to hydrophobic after sample preparation can be influenced by wrapping paper serving as a donor for hydrocarbons, which it has most likely adsorbed itself through contact with atmospheres rich in adventitious carbon before [24]. In Fig. 2 a), we present a comparison of samples stored at ambient conditions with samples wrapped in woodfree tissue. As Fig. 2 a) shows, the wrapped samples depict a steady hydrophobic SCA, while samples without the paper directly exposed to the surrounding atmosphere transition back to a more hydrophilic behavior once humidity rises. While water molecules present at the surface in initial stages of aging can support the adsorption of volatile carbonous species and therefore the transition from hydrophilic to hydrophobic SCAs, water molecules present at the surface after the hydroxylation process might form hydrogen bonds and prevent further adsorption of nonpolar groups [25,34–38]. We assume that the formation of such hydrogen bonds on the samples stored in increased humidity without the wrapping paper are responsible for the decrease in SCA and its subsequent stagnation. These results confirm the need of a strict control of the storage conditions. Fürbacher et al. [34] as well witnessed a delay in the expected wetting transition from hydrophilic to hydrophobic upon sample exposure to the atmosphere. While vacuum storage [37,38] or annealing [26,39] can be conducted in order to achieve a quick wetting transition to hydrophobic behavior, we refrain from this method due to possible side contamination and unsteady atmospheres depending on previous furnace/chamber use [34]. To investigate stable wetting states, we decided to focus on aged samples in this study as the wetting behavior of freshly prepared surfaces might be influenced by metastable states and intrinsic material properties like microstructure [20]. In ref. [20], a saturation of adsorbed hydrocarbons could not be observed. Therefore, here we extended the storage time to investigate if a saturation is reached after prolonged times. A measurement series with mechanically polished copper surfaces packed in woodfree tissue for 3, 6, and 43 weeks was conducted with combined SCA and XPS measurements on identically stored and handled comparative samples (Fig. 2 b)).

During aging, the samples show a strong increase of the elemental concentration ratio of carbon to copper (C/Cu), a common parameter to characterize the degree of carbon coverage [18,26]. Additionally, the measurements demonstrate that the detected carbon is not bound as graphite or in form of pure hydrocarbon chains, but rather with the participation of oxygen, forming functional groups with polar character, like hydroxyls (–OH), aldehydes (–COH), ketones (C═O), ethers (–O-), carboxylic acids (–COOH), or esters (–COOC-). A second parameter was introduced, the polarity ratio of the adsorbate layer. For reasons of simplicity, the polar groups in which the carbon is bound to one or two oxygen atoms are summarized. The polarity ratio of the adsorbate layer ($C_{pol}/C_{nonpol}$) – a relevant parameter for the wetting behavior [40,41] – is then calculated as the quotient of the proportion of carbon bound in these polar binding situations to carbon bound in aliphatic hydrocarbon chains. Besides the fluctuation after 6 weeks in Fig. 2, the polarity ratio also decreases with time in agreement with ref. [20]. Despite the fluctuations in the polarity ratio occur and the amount of accumulated carbon groups differs strongly over time, the SCA measurements of the three samples under investigation depict stable wetting conditions and prove the independence of differences in the adsorbate layer once a contamination film of sufficient thickness or density is adsorbed.

In order to gain insight into the deep structuring of the adsorbate layer, a XPS tilting experiment was conducted. To compare the elemental concentrations measured at two different tilting angles (30° and 90°) relative to the sample surface, in equation (4) their contrast is calculated. It serves as an estimation of the element depth location, because with decreasing tilt angle the information depth is reduced.

$$Contrast = 100*\frac{C(30^{\circ}) - C(90^{\circ})}{C(30^{\circ}) + C(90^{\circ})} \quad (4)$$

As expected from the literature [23,26,42], Fig. 3 shows that the adsorbed nonpolar hydrocarbon groups form the outermost layer of the samples with oxygen and copper underneath. Analysis of the C1s core level spectrum indicates that the nonpolar HCH groups point outwards and therewith form the primary contact to applied water droplets. The polar groups are deeper and seem to be responsible for the adhesion of the adsorbate to the metal or rather oxide surface. This may be an explanation why in Fig. 2 b) all samples show a similar SCA, although the ratio C/Cu as well as $C_{pol}/C_{nonpol}$ change. The hydrocarbon coverage ratio and layer thickness/density seem to be sufficient to dominate the water-substrate interaction with a nonpolar and, therewith, hydrophobic termination of the adsorption layer.

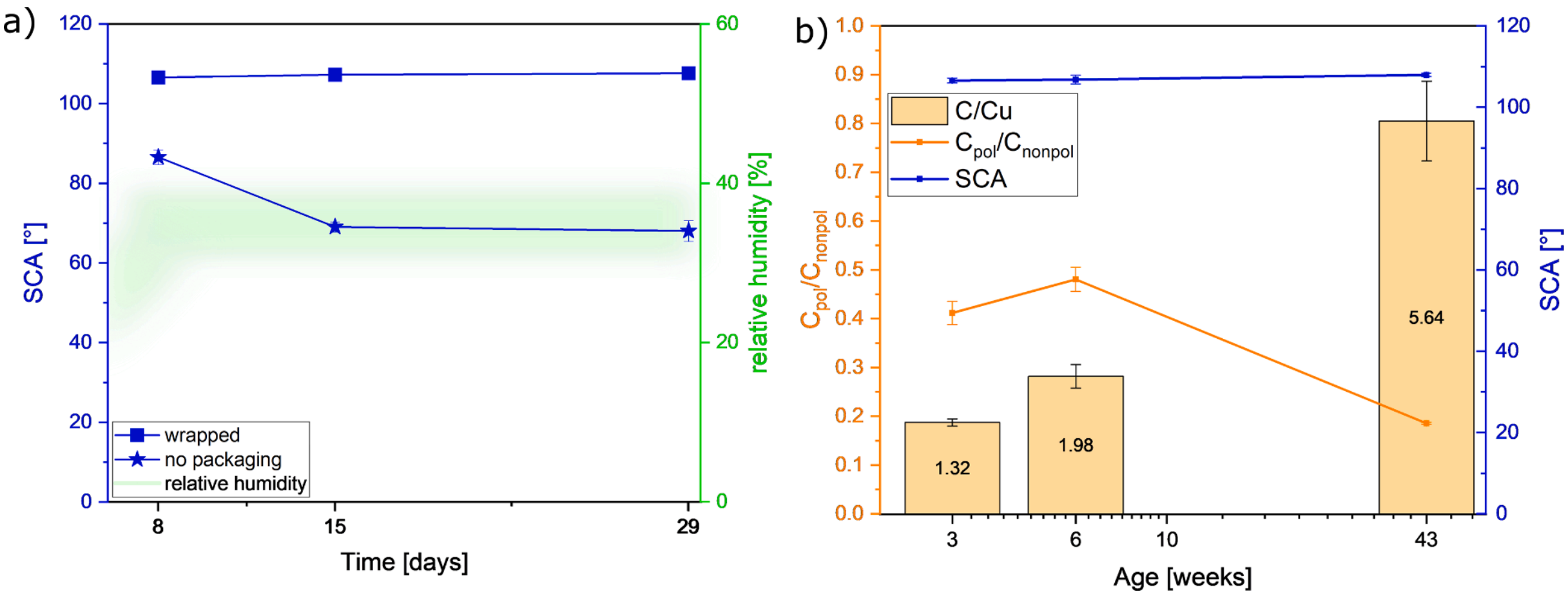

**Fig. 2.** Aging behavior of mechanically polished copper samples. (a) Mean value of measured SCAs with standard deviation and an indication for the relative humidity over storage time. Comparison of samples wrapped in woodfree tissue with samples stored in ambient conditions without any packaging. SCA measurement conditions: 21–23 °C; RH.: 29 % (day 8), 40–44 % (day 15 and 29); at least 9 droplets analyzed for each data point. (b) Mean value of measured SCAs with standard deviation (blue line) and mean $C_{pol}/C_{nonpol}$ (orange line) as well as mean C/Cu (orange bars) with standard deviation. Samples were wrapped in woodfree tissue. The SCA is stabilized while changes in the surface chemistry still occur. SCA measurement conditions: 21–23 °C; 35–40 % RH.; at least 7 droplets analyzed for each data point.

These results show that once an adsorption layer of sufficient thickness or density is grown, a further accumulation and even slight fluctuations in the polarity ratio do not significantly affect the SCA. Korczeniewski et al. [21] conducted simulations showing that the strongest change in SCA is observed after formation of a monolayer of hydrocarbons on copper single crystals, whereas additional layer growth only shows little effect, which is in good agreement with our experimental observations regarding C/Cu. This supports our results from an earlier work [20], indicating that there might be a material and microstructure-dependent threshold concentration of C/Cu which induces a wetting transition from hydrophilic to hydrophobic. This threshold might also depend on hydrocarbon type [43]. Our results suggest that the adsorption is not solely chemically driven but that after chemisorption, physisorption of hydrocarbons continues as the latter is not characterized by a self-limiting effect [44]. We assume that chemisorption is responsible for the wetting transition from hydrophilic to hydrophobic, while subsequent physisorption causes further gradual increase in SCA until the interaction forces are fully dominated by the hydrocarbon layer. This could also explain why we saw a levelling of copper SCA around 100° in a previous study [20] observing the wetting transition due to chemisorption, while samples in Fig. 2 level around 108° due to subsequent physisorption. However, the process of hydrocarbon adsorption is highly complex and elevated theoretical models will be needed especially for the investigation of multilayer hydrocarbon adsorption [43].

**B) Influence of oxide chemistry**

DLIP structures are generated by complex chemical as well as topographical modifications. It is known in the literature, that during the laser patterning of copper, both CuO and $Cu_2O$ are formed, though $Cu_2O$ seems to dominate the oxidic surface [29,45]. To separate the influence of topography and chemistry, we prepared flat surfaces with the two different oxides and pure metallic copper in a first step by sputter deposition. LSM measurements depict a similar roughness $S_a$ of 3–4 nm for the sputtered surfaces which is in the range of the roughness of the polished substrates. Table 1 summarizes the roughness characterization and shows the S-ratio to be 1.000 for all surfaces within the resolution of the LSM suggesting a negligible influence of topography on the wetting behavior. FIB cross-sections displayed in Fig. 4 d) – f) allow for the measurement of the thickness of the deposited layers. To enhance the contrast between the copper substrate and the deposited layers and to avoid rounding effects of the cutting edges during FIB preparation, a $Au_{80}Pd_{20}$ layer was deposited before preparing the FIB cross-sections. Table 1 shows the layer thickness ranging between 130 nm and 147 nm. A thickness above 100 nm was pursued to assure a negligible influence of the underlying copper substrate in the water interaction. XRD measurements were performed to ensure the growth of the desired oxides. The star quality reference patterns shown as vertical lines in Fig. 4 d) are slightly shifted for the two oxide types compared to the peaks in the recorded spectra. Such peak shifts were also observed in other studies reporting the sputter deposition of oxidic copper layers [46]. This might depend on the oxygen pressure [47] and is an indication for deviations in the lattice constants [48,49], most likely caused by point defects and non-stoichiometry of the $Cu_2O$ phase.

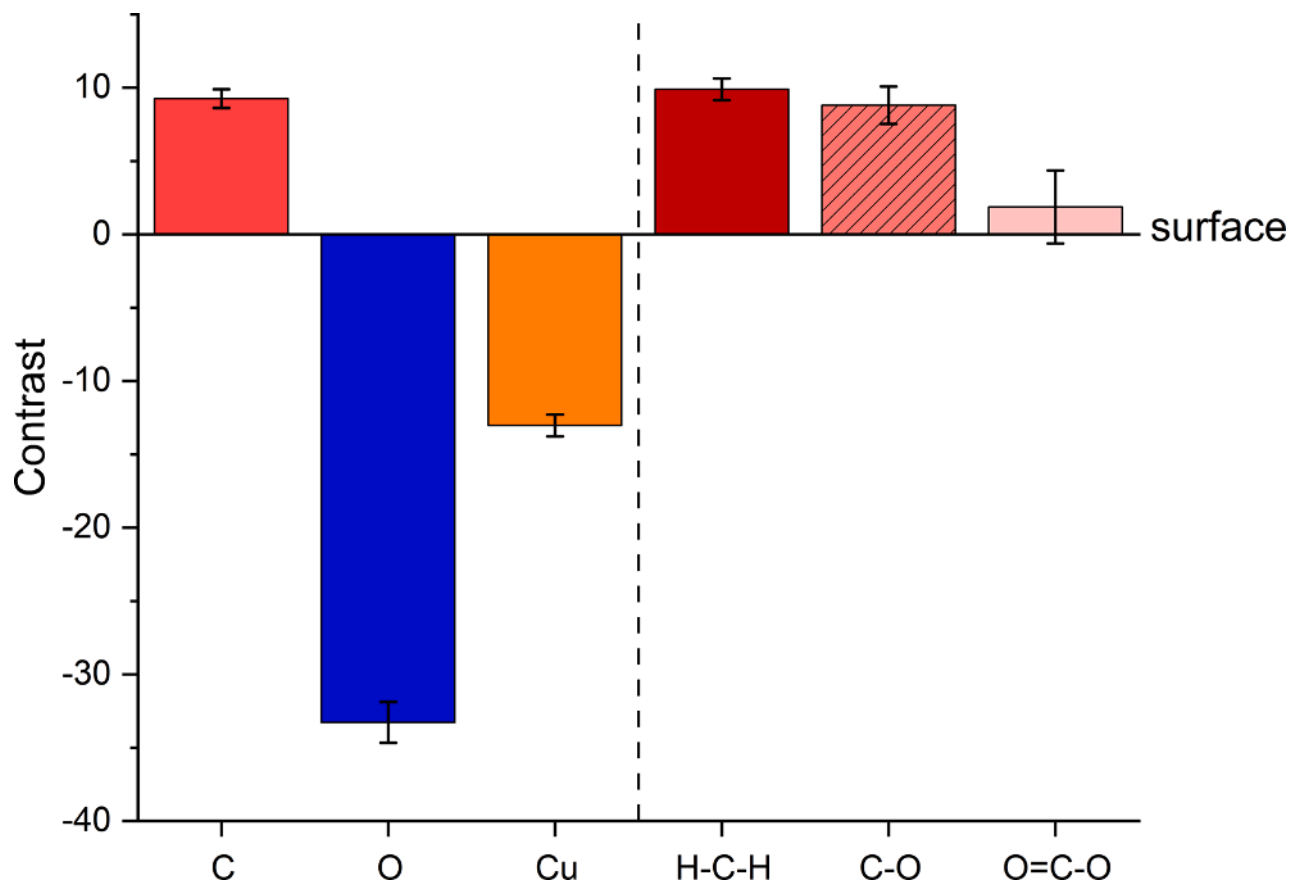

**Fig. 3.** Contrast calculated according to equation (4) from XPS measurements perpendicular (90°) and under an angle of 30° to the sample surface indicating the depth distribution of the measured elements on wrapped samples stored for 57 weeks.

The analyses results from Fig. 4 and Table 1 confirm the production of flat surfaces differing in copper oxide states. After aging the samples, wetting as well as near surface chemical characterization were conducted. Instead of using the same samples for both analyses, only identical treated samples were used to avoid interactions respectively cross-contaminations between samples used for SCA and XPS analysis. Fig. 4 e) summarizes the SCAs measured 10 s, 60 s and 180 s after deposition of 3 µl water droplets. The results reveal, that for each surface type, after 10 s a stable wetting state is already reached and no significant droplet spreading can be observed within three minutes after droplet deposition. The CuO samples show a very slight anisotropy of approximately 2° which could be induced by some preparation artefacts from the underlaying polished substrate but is not expected to be a result of the sputter-deposition. When averaged over the two measurement directions, after 60 s, no significant differences can be observed between the sample types as they all range between 111° and 112° suggesting a negligible influence of copper oxide state on the water SCA.

Park et al. [49] found an increase in SCA on magnetron sputtered copper oxide thin films with increasing annealing temperature, which proved to not affect the oxide composition (they found a mix of amorphous $Cu_2O$ and monoclinic CuO) but to enhance surface roughness. They attributed the increase in SCA to a lotus effect, but no details were provided on sample age or storage and the droplets with a very small volume of 0.2 µl and corresponding short contact lines might be easily disturbed by certain roughness features. Ghotbi et al. [50] prepared $SiO_2$/Cu and $SiO_2$/CuO thin films by hydrothermalization varying in coating duration and strongly in surface morphology. They observed hydrophilic behavior with SCAs < 40° for all oxide films but also did not give details on the storage and measurement conditions. Zhao et al. [26] employed the hydrothermalization method to form films consisting of CuO and $Cu_2O$ polyhedrons which both showed superhydrophobic properties after storage in C-rich atmospheres. XPS analyses revealed the adsorption of nonpolar hydrocarbon chains and theoretical calculations supported the idea of a strong chemisorption of carboxylic acid. They found that oxide composition does not influence the superhydrophobicity which is controlled by strong chemisorption of volatile carbon species, but the storage time varied greatly for the two oxide types. Our results go along with the observations made by Zhao et al. [26] as oxide composition does not seem to significantly affect the long-term SCAs on the sputter-deposited flat surfaces.

**Table 1**
Sputter layer thickness and roughness parameters determined by LSM for the sputter-deposited layers.

| | Cu | CuO | $Cu_2O$ |
|---|---|---|---|
| Layer thickness (FIB) [nm] | 147 ± 14 | 140 ± 7 | 130 ± 5 |
| Sa [nm] | 4 ± 0 | 4 ± 1 | 3 ± 1 |
| S-ratio | 1.000 ± 0.000 | 1.000 ± 0.000 | 1.000 ± 0.000 |

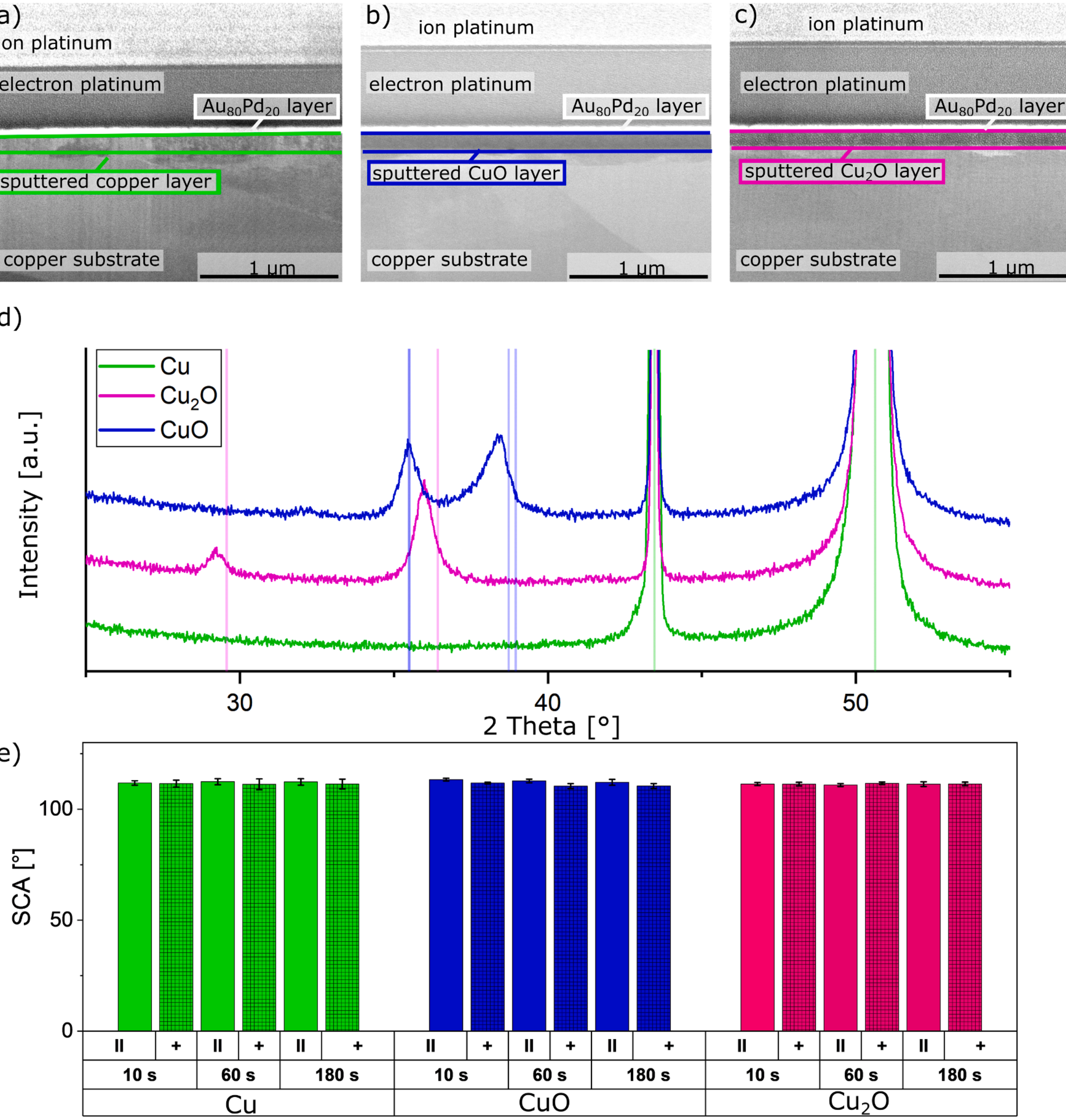

**Fig. 4.** FIB cross-sections of the sputtered surfaces coated with $Au_{80}Pd_{20}$ to enhance the contrast to the sputtered Cu layer (a), sputtered CuO layer (b) and $Cu_2O$ layer (c). (d) XRD-analysis of the sputtered surfaces confirming the formation of Cu, CuO and $Cu_2O$. Vertical lines represent the reference patterns (Cu: 04-009-2090, CuO: 00-045-0937, $Cu_2O$: 04-007-9767). (e) Mean value of SCAs parallel (II) and perpendicular (+) to the sample axis with the standard deviation of aged sputter-deposited Cu, CuO and $Cu_2O$ samples measured 10 s as well as 60 s and 180 s after droplet deposition. SCA measurement conditions: ≈ 22°C; 39–42% RH.; at least 10 droplets analyzed on 3 samples per type for 10 s measurement; at least 7 droplets analyzed on 3 samples per type for 60/180 s measurement.

Near surface chemical characterization by XPS analysis was conducted to further verify the oxide state at the uppermost surface layer. XPS characterization of copper and its oxides is prone to large uncertainties due to rather small chemical shifts of the Cu 2p photoelectron line and strong overlaps (compare Fig. 5). Due to these difficulties in reliable copper oxide characterization by XPS, additional reference samples sputtered together with the samples from the analysis in Fig. 4 were used to obtain model spectra for the different oxide types (see Fig. 5) and used in all subsequent XPS analysis of Cu 2p and CuLMM for deconvolution to identify the contributions of different oxidation states (see supplementary material for details).

While the Cu 2p photoelectron line of CuO shows a just observable chemical shift (< 1eV) to metallic copper and a rather significant strong satellite structure at about 940 eV ($Cu^{2+}$) as identification criteria, $Cu_2O$ and metallic copper are nearly not distinguishable in the Cu 2p region [51]. Also, the difference of O1s chemical shifts (< 0.5 eV) is too small to serve as criteria. The Cu LMM Auger lines in contrast show rather significant structures, for all three binding states. This offers the possibility of an identification by the peak shape of the Auger lines. Fig. 6 summarizes the elemental compositions retrieved from the Cu LMM, the Cu 2p and the O1s spectra of the aged sputter-deposited samples. Considering that the differences in peak shapes for Cu 2p are not very significant these results are only given for the sake of completeness. The same holds true for the O1s core level spectra. So as expected, the results based on analysis of Cu LMM, Cu 2p and O1s show some deviations, but all tend to confirm the identification of the sputter-deposited oxides as dominated by CuO and $Cu_2O$, respectively. The Cu-treated sample reveals native oxidation after aging which was not detected by XRD in Fig. 4. The quantitatively most reliable results are based on the Cu LMM analysis.

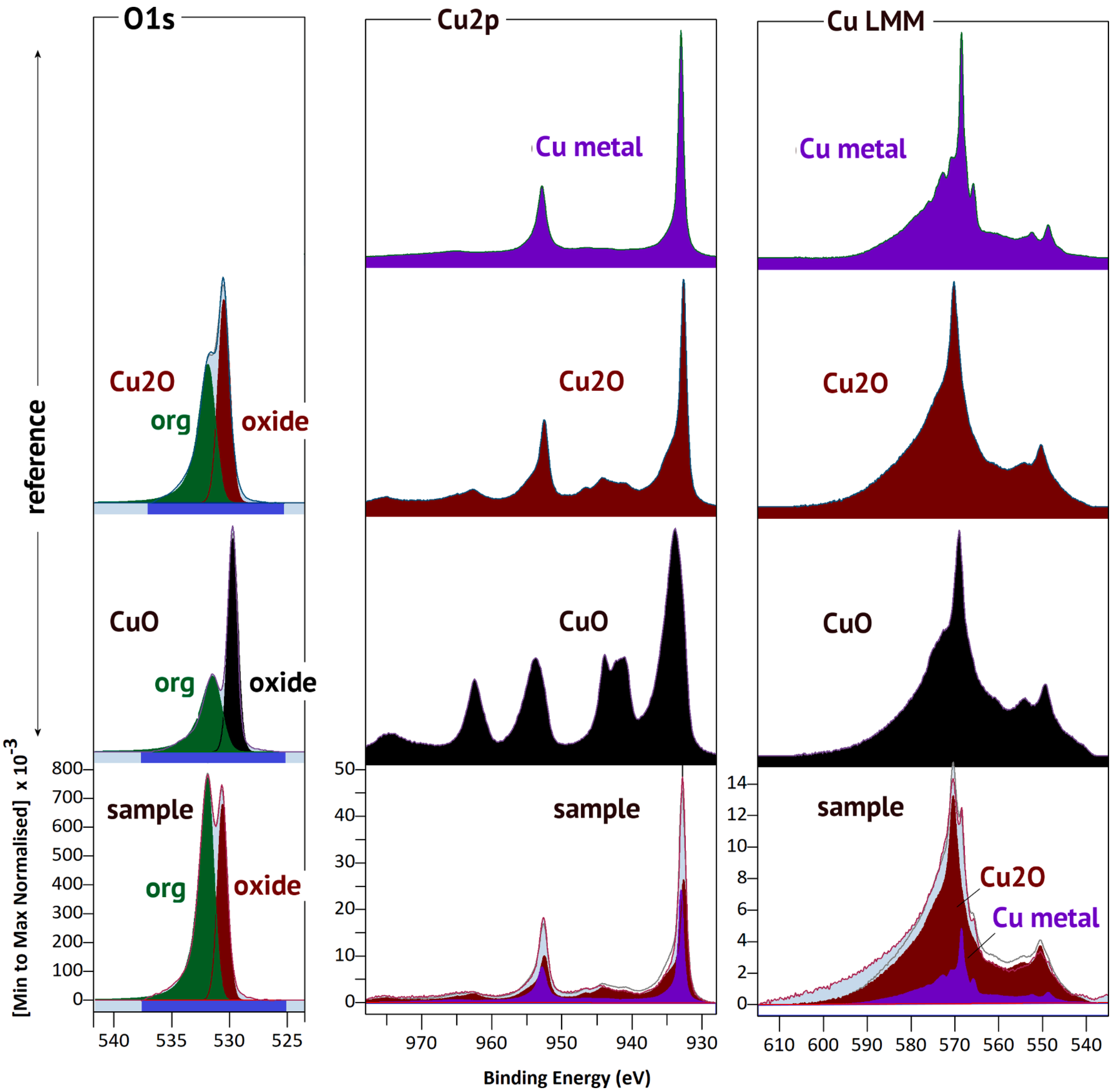

**Fig. 5.** Comparison of O1s, Cu2p and Cu LMM core spectra for reference surfaces of Cu (sputter cleaned bulk sample), $Cu_2O$ respectively CuO (sputter-deposited). In the lower line the respective spectra for a polished aged Cu sample are shown inclusive a deconvolution into the distinguished distributions.

Our XPS analysis reveals negligible sulfur and nitrogen contaminations on the aged samples in the overview spectra (Fig. 7 a)). The detected carbon is predominantly bound as aliphatic hydrocarbon, and confirms the interpretation of omnipresent adventitious carbon (compare section A of the manuscript), whose nature is surprisingly still discussed [52]. Detail C1s spectra (Fig. 7b) show that besides the dominant portion of carbon bound in aliphatic hydrocarbons a significantly smaller portion has one or two oxygen atoms as direct binding partners. This suggests the involvement of organic compounds such as ethers, aldehydes, ketones, esters, alcohols, carboxylic acids, which are airborne and under typical atmospheric conditions omnipresent. These oxygen-containing organic species are subsumed and abbreviated here as OCO (carbon with two oxygen as binding partners in an aliphatic hydrocarbon chain) and HCO (carbon with one oxygen as binding partner in an aliphatic hydrocarbon chain) and form a small part of the adsorption film. The ratio $C_{pol}/C_{nonpol}$ is similar for the sample types with 0.31 $\pm$ 0.02 for Cu compared to 0.28 $\pm$ 0.01 for CuO and 0.29 $\pm$ 0.03 for $Cu_2O$, while the Cu samples show a greater uptake of volatile carbons. As the Cu sample is not passivated once it comes in contact with the atmosphere after the sputter deposition, it is likely chemically more reactive than the two passivated oxidic surfaces. In a previous study [20], chemically more active samples due to an increased surface defect density showed an enhanced uptake of hydrocarbons. We believe that surfaces with a higher affinity to oxygen like the Cu sputtered samples might also depict a higher affinity to hydrocarbons. Besides, interaction with water vapor influencing hydroxylation of the surfaces might be more pronounced for the pure copper surface than for the oxidic samples (see section A for discussion of water vapor interaction).

It must be noted that for our deposited copper films, adsorbed water molecules do not only influence hydroxylation but also native oxidation processes that can be drawn responsible for the detection of mainly $Cu_2O$ on the Cu-treated samples (compare Fig. 6a) [53]. In order to evaluate the mechanisms of hydrocarbon adsorption on the three different sample types, high resolution analysis in early-stage aging would be necessary but was out of scope of this study as our main goal was to determine whether differently passivated copper surfaces depict varying SCAs after stable states were reached due to pronounced adsorption film formation.

Independent on oxide state, all three sample types build strong adsorption layers dominated by nonpolar hydrocarbons, which ultimately determine and dominate the wetting properties. The resulting hydrocarbon adsorbate film masks the surface and shields the metal oxide underneath. The results from Section A) and B) allow for a separation of chemical and topographical influences on laser patterned surfaces. Before investigating complex hierarchical patterns, in a next step, the influence of a random topography paired with the oxidic surface was investigated.

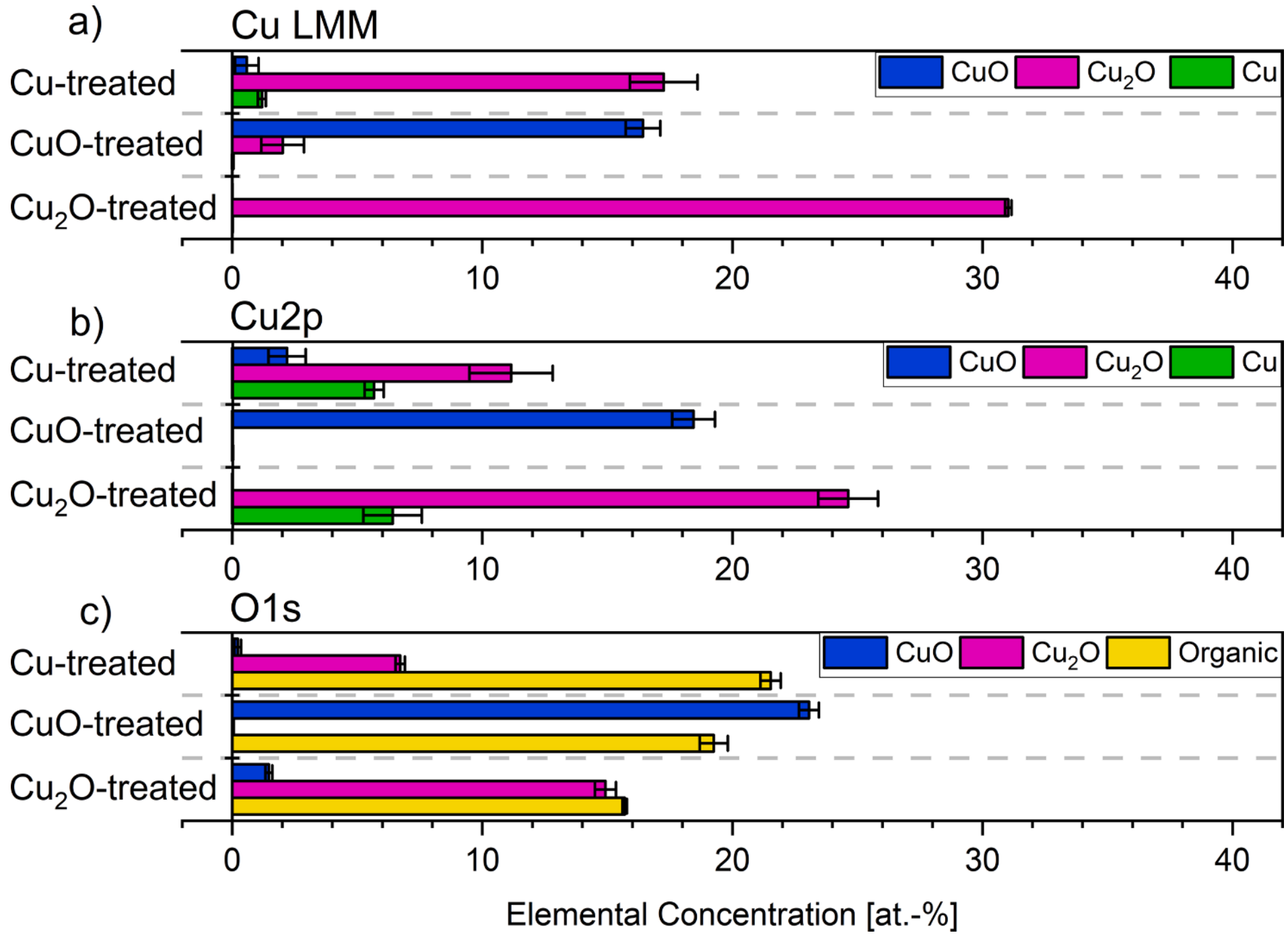

**Fig. 6.** Elemental concentrations in at.-% obtained from the XPS detail spectra Cu LMM (a), Cu2p (b) and O1s (c). The sputter treatments are specified on the y-axis, the calculated concentrations are represented by colored bars.

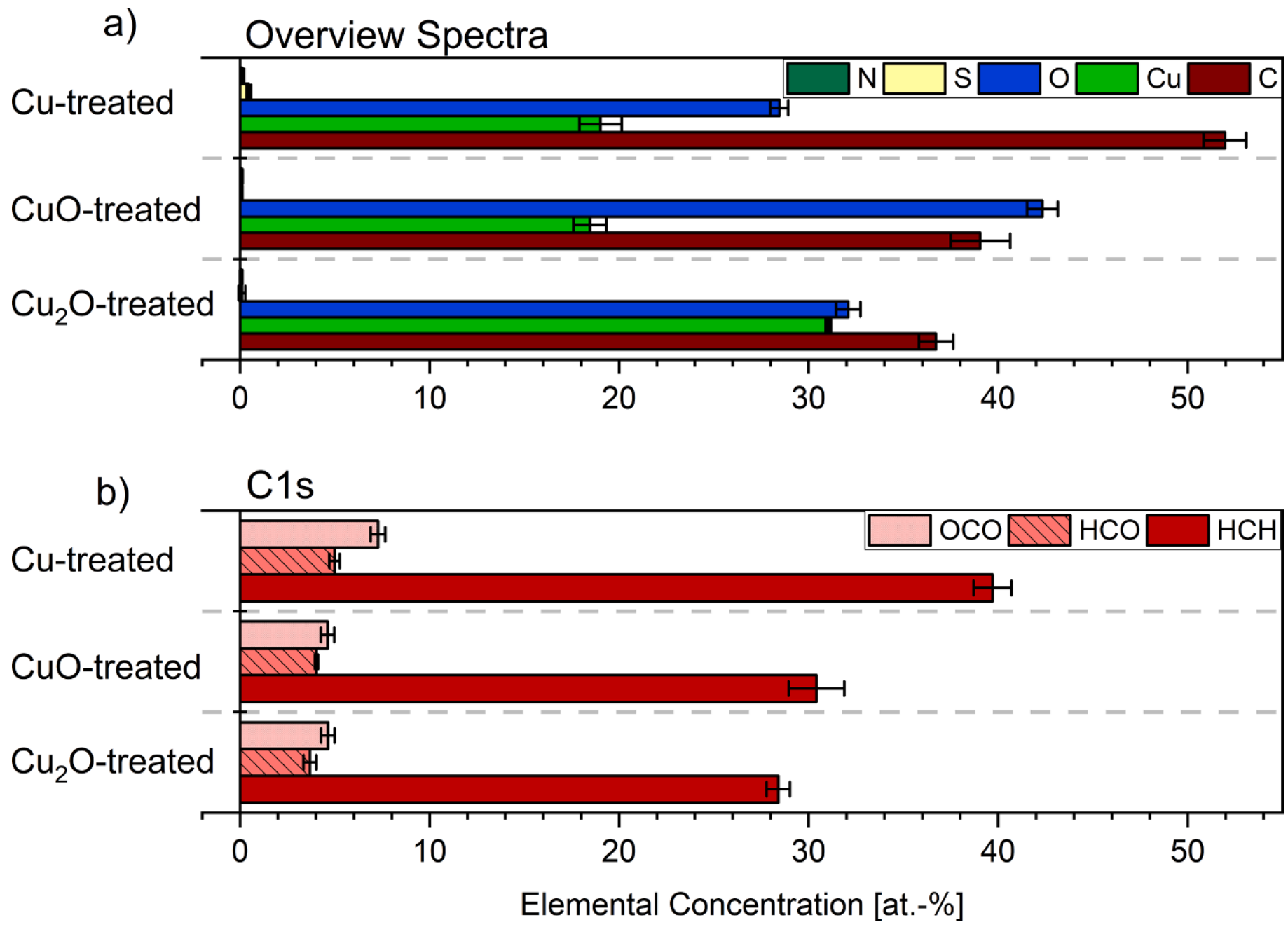

**Fig. 7.** Elemental concentrations calculated from XPS overview spectra (a) for all detected elements and (b) for carbon, broken down by its binding states according to the deconvolution of C1s core level spectra, of the sputtered samples. Sample sputter treatments are given on the y-axis.

**C) Influence of topography**

LR samples were produced by laser processing of polished copper samples with one ps beam and parameters were chosen in order to minimize directional topography following the laser scanning direction. Fig. 8 summarizes the topographical characterization conducted by SEM and FIB analysis, AFM, and LSM measurements.

The LR-samples show a random roughness consisting of melt structures and melt droplets varying in size forming undercuts and pores. These features complicated the AFM analysis in Fig. 8 c). The AFM tip is a non-equal-sided pyramid. The minimum angle of inclination is 65°. Thus, the maximum angle of the pyramidal shape with respect to its axis is less than 25°. Therefore, no feature with a higher aspect ratio, especially undercuts, can be characterized by AFM. Due to the irregularities of the surface, a slow scan rate was necessary. The LSM analysis in Fig. 8 f) does not depict significant differences in topography perpendicular and parallel to the laser scanning direction. Chemical characterization by XRD revealed the oxidic character of the melt features with a dominance of $Cu_2O$ over CuO (compare Figure S1). Table 2 summarizes the roughness parameters obtained by LSM and AFM. The average surface roughness is 20 nm as measured by LSM, which is larger than that of the polished samples. The increase in surface area determined by the S-ratio differs greatly between the two different measurement techniques. Due to the lower lateral resolution of the LSM, the measured S-ratio is approximately 22 % lower than that measured by AFM.

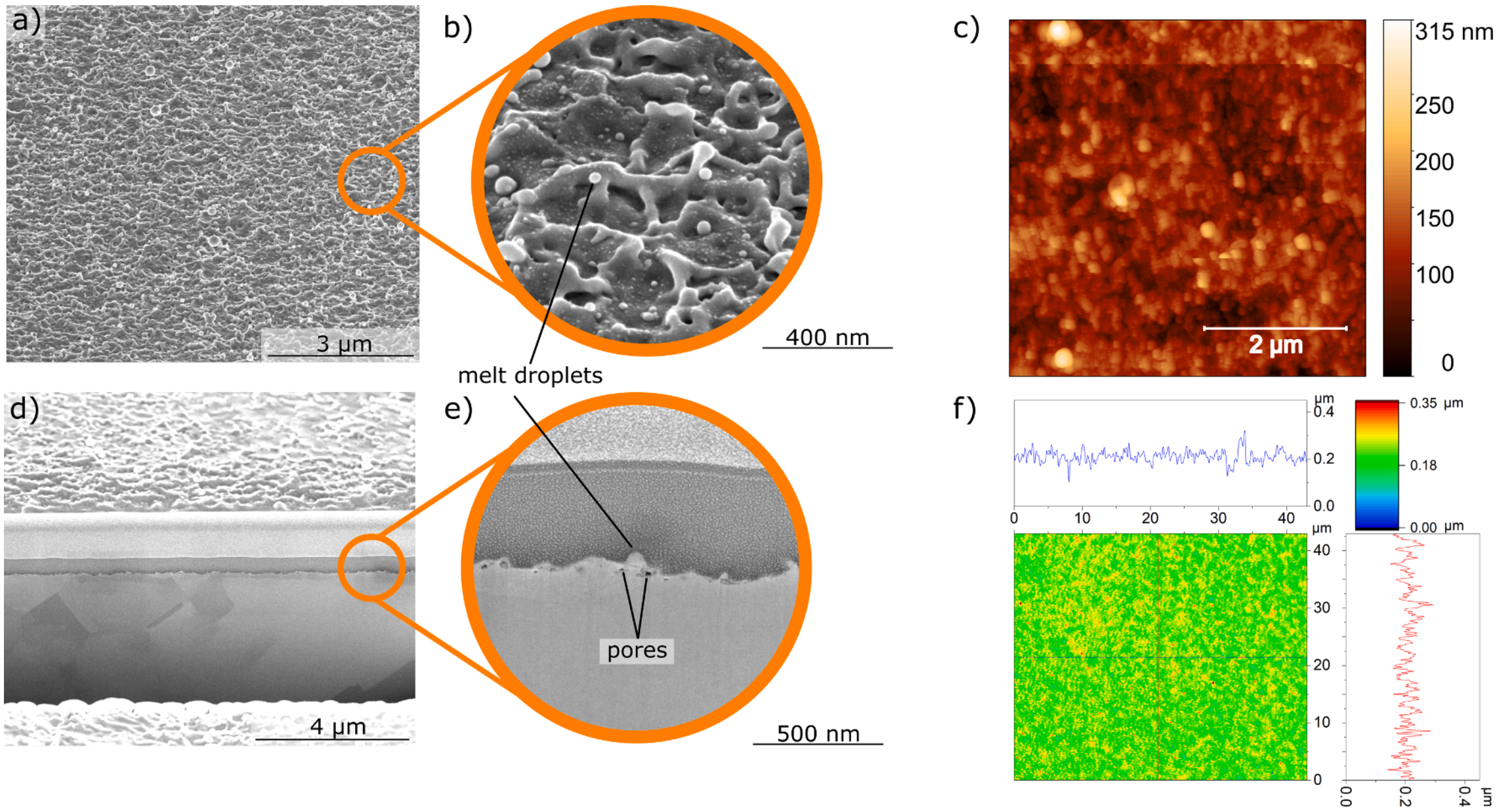

**Fig. 8.** Topographical characterization of a LR-Sample. (a) SEM-image of the 52° tilted surface with a magnified image in (b). (c) AFM characterization of the surface. (d) FIB cross-section with a magnified image in (e). (f) LSM characterization with height profiles parallel and perpendicular to the laser scanning direction.

Larger scale XPS analysis, that does not resolve the melt structures and melt droplets was conducted to investigate whether the picosecond laser treatment influenced the adsorption of volatile hydrocarbons with different functional groups. The analysis of the overview spectrum in Fig. 9 a) shows a clear dominance of copper, oxygen, and carbon as expected for aged laser treated copper surfaces. The detailed analysis of the C1s core level spectra (compare Fig. 9 b)) reveals, similarly to the polished and sputtered samples, a strong adsorption of hydrocarbons and considerably less oxygen-containing adventitious carbon groups leading to a polarity ratio ($C_{pol}/C_{nonpol}$) of 0.28 ± 0.1.

Following our argumentation, that Cu LMM spectra are the most reliable source for oxide characterization, a clear dominance of $Cu_2O$ follows. Some studies report an initial dominance of CuO right after laser processing which converts into $Cu_2O$ upon exposure to oxygen during sample storage [31]. Cheng et al. [31] also attributed the witnessed wetting transition to the deoxidation of CuO to $Cu_2O$. Our results contradict this hypothesis as we found both oxide types to be hydrophobic upon exposure to ambient air due to hydrocarbon adsorption. Gaudiuso et al. [39] observed both oxide states together with $Cu(OH)_2$ on femtosecond laser structured and low temperature heat treated surfaces but attributed the wetting transition to the adsorption of hydrocarbons on the hierarchical topography which was accelerated by the heat treatment. Hydrocarbons also proved in our study to be the relevant factor for the latter discussed transition in the wetting behavior. The O1s spectrum in Fig. 9 c) reveals that most of the immediate surface oxygen detected by XPS is bound in organic species with the remaining oxygen being bound preferably in $Cu_2O$ (see Cu LMM).

**Table 2**
Roughness parameters of the LR-Samples determined by LSM and AFM.

| | LR-Sample |
|---|---|
| Sa [nm] | 20 ± 0 |
| S-ratio AFM | 1.307 ± 0.07 |
| S-ratio | 1.025 ± 0.001 |

Fig. 10 shows the results of SCA measurements with the standard deviation on the LR-samples perpendicular and parallel to the laser scanning direction. The aged surfaces show a hydrophobic behavior and a slight anisotropy of the SCA appears with mildly increased angles perpendicular to the laser scanning direction (increase of approximately 2° after 10 s and of 3° after 60 s / 180 s). Though the topographic analysis in Fig. 8 does not show a visible structural anisotropy, the SCA reveals a weak direction dependency indicating that the topography has a very slight variation parallel and perpendicular to the laser scanning direction, which is below the resolution of the LSM analysis in Fig. 8. Though a parameter study was carried out to find the laser parameters, which deliver the most homogeneous pattern, the roughness can be overlayed with the trace of the single laser beam that was scanned over the samples multiple times with a hatch distance of 13.5 µm resulting in the observed wetting anisotropy. Also, the SCA parallel to the laser scanning direction shows a slight decrease with time suggesting potential capillary forces along the scanning direction of the laser beam. Still with the wetting anisotropy being in the magnitude of the standard deviation of the SCAs measured parallel, it can be considered as very mild.

As a reference, copper samples were polished, packaged, and measured on the same day as the LR-samples and depicted a SCA of 108° ± 2, which is in good agreement with the results from Fig. 2. The minor roughness induced by the picosecond LR process increases the SCA by approximately 25 %. This is in agreement with the literature, where laser-induced surface features lead to a hydrophobization [28,30,54,55].

The S-ratios shown in Table 2 were used to calculate the theoretical SCA according to Wenzel [15], which underestimated the measured SCA greatly, indicating that either even the fine AFM tip was not able to capture enough details in order to obtain the correct increase in surface area or the droplet forms air-inclusions with the surface depicting a wetting behavior according to Cassie and Baxter [16]. For the calculation of the CB contact angle, the wetted fraction of the rough surface needs to be determined which poses a great challenge as this value is not accessible via direct measurement. Moreover, transition states between CB and Wenzel exist [56]. Though the models from Wenzel and from Cassie and Baxter are very well established, it is known in the literature that there are wetting states that cannot be represented by them as not all topographic features are expected to influence the resulting wettability or show the potential to trap air [57]. According to equation (1) a r factor of 2.3 would be needed to approximate the measured SCAs around 135° for a Wenzel wetting state. As our previous investigations showed that a carbonous adsorption layer dominates the SCA over differences in the underlaying oxide state, clearly topography can be held responsible for the increase in SCA. It must be mentioned though that the topography can only show its full potential after this adsorption took place. The production and detailed analysis of the sputter-deposited oxides as well as the LR-samples serve as a basis for the comprehension of complex wetting phenomena as observed on anisotropic hierarchical structures produced by DLIP. DLIP surfaces are not only characterized by the main periodic pattern, but also portray secondary roughness (sub-pattern) e.g. as LIPSS, craters, and/or a decoration with melt structures or even smaller nanoparticles [45,58,59].

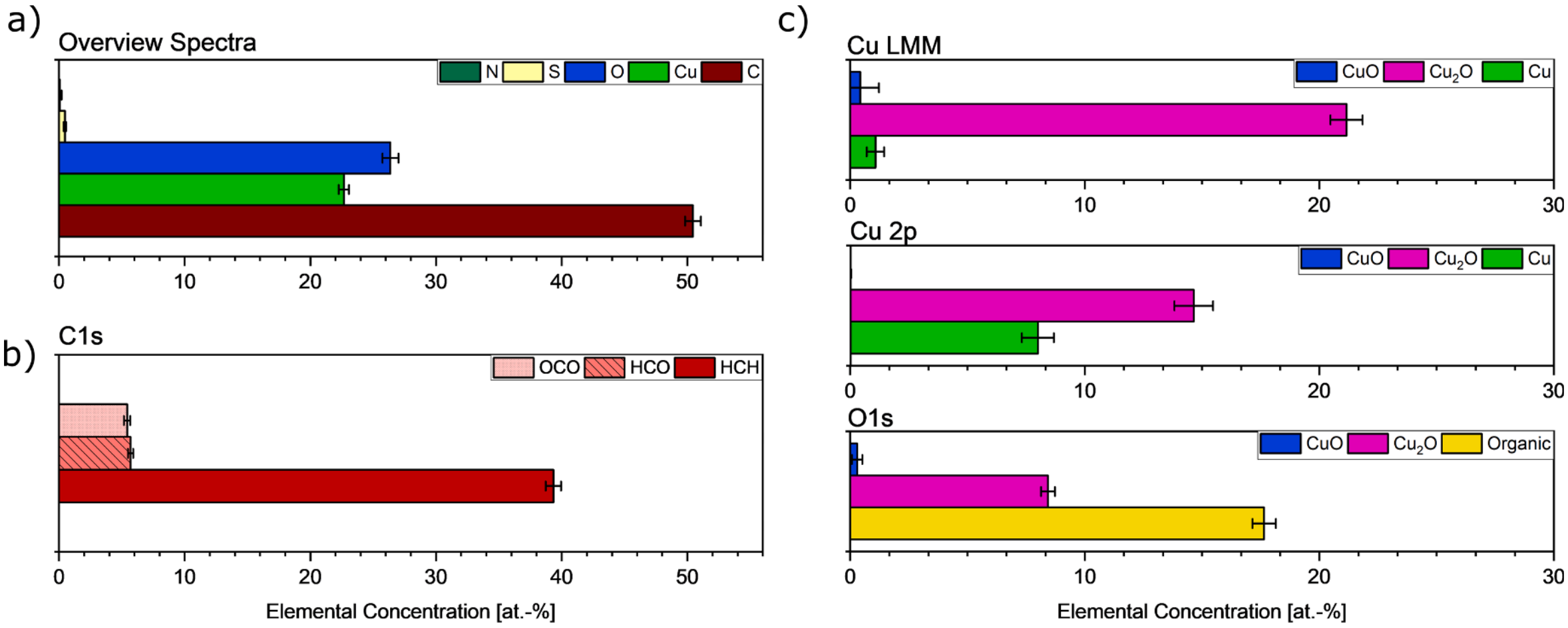

**Fig. 9.** Results of the XPS-analysis of the LR-Samples. (a) elemental concentrations from overview spectra, and (b) for carbon, broken down to its binding states according C1s core level spectra, as well as (c) for copper and oxygen from LMM Auger, respectively Cu 2p and O1s detail spectra.

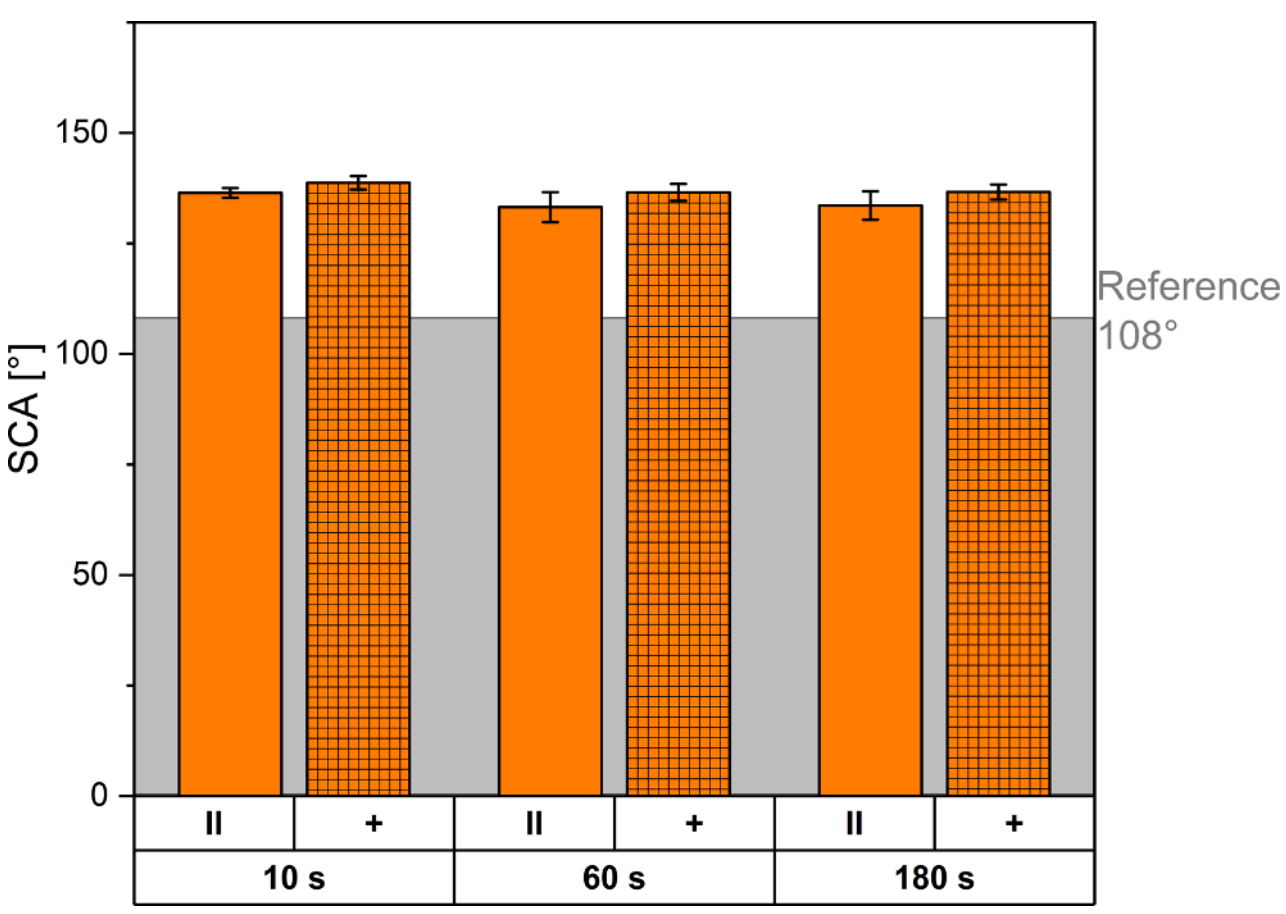

**Fig. 10.** Mean value of SCA measurements on LR-samples 10 s, 60 s and 180 s after droplet deposition parallel and perpendicular to the laser scanning direction with standard deviation. Only a very mild anistropy is visible. Measurement conditions: ≈ 23 °C; ≈ 39 % RH.; 15 droplets on 3 samples for 10 s measurement; 6 droplets on 3 samples for 60/180 s measurement.

**D) Complex DLIP patterns**

While copper oxide state does not seem to significantly alter the wetting response after adsorption of hydrocarbons, topography showed a pronounced effect even for simple structures. To determine the influence of a complex surface topography in more detail, different DLIP patterns were applied to polished copper surfaces. The structure periodicity was set to 6 µm, while the structure depth and therewith the aspect ratio was varied by laser power adaption of the picosecond laser. To fabricate a sample with changes in the nanoroughness with an identical primary line-like pattern, additionally a femtosecond laser was used to duplicate one of the pico-second laser structures. This does not only allow for the detailed analysis of the wetting response to slight topographic changes but also the characterization of surface morphology and chemistry induced by different pulse durations.

Fig. 11 summarizes the topographical characterization of the four sample types. Numbers in the sample names indicate the pattern depth of 500 nm for the fs-DLIP sample (named F500) and the ps-DLIP sample (P500) or 1000 nm respectively (ps-DLIP sample P1000). The LSM-scans in Fig. 11 a) – c) are combined with profile plots parallel and perpendicular to the main pattern. All three sample types depict periodic line patterns spreading homogeneously over the scanned area with the P1000 samples being approximately twice as deep as the other two sample types. The mean height of the profile elements $R_c$ was measured in 9 different sample areas and results were averaged in Table 3. The desired structure depths were produced without significant differences between the F500 and the P500 samples. Also, the average roughness $S_a$ calculated in three different sample areas, depicts comparable values for the F500 and P500 samples (147 and 134 nm, respectively) and an increased roughness of 272 nm for the P1000 samples. It must be noted that the LSM measurement cannot resolve all structural details due to its limited resolution (120 nm lateral, 10 nm in height). AFM scans in Fig. 11 which also have a limited resolution show an increased roughness in the structural valleys of F500, which might contribute to the slightly higher $S_a$ values while the decoration with nanoparticles on the peaks of the P500 sample might not be fully detected by the LSM. Clearly, the pile-up of particles on the P1000 sample was detected though (compare Fig. 11 c). For the increase in surface area (S-ratio) the same trend can be observed. Both ps samples show more oxidic particles on the peaks and especially for the P1000 sample, the AFM scan reveals melt droplets agglomerated on the maxima of the pattern. The LSM profile scan along such a structural maximum also reveals the more pronounced roughness. The fs sample on the other hand is clearly substantially rougher inside the valleys. On the AFM scan of the polished reference sample, only a few scratches from the final polishing step are visible leading to a roughness of 4 nm. For a more detailed analysis of the topographic features, SEM and FIB characterization were carried out (compare Fig. 12).

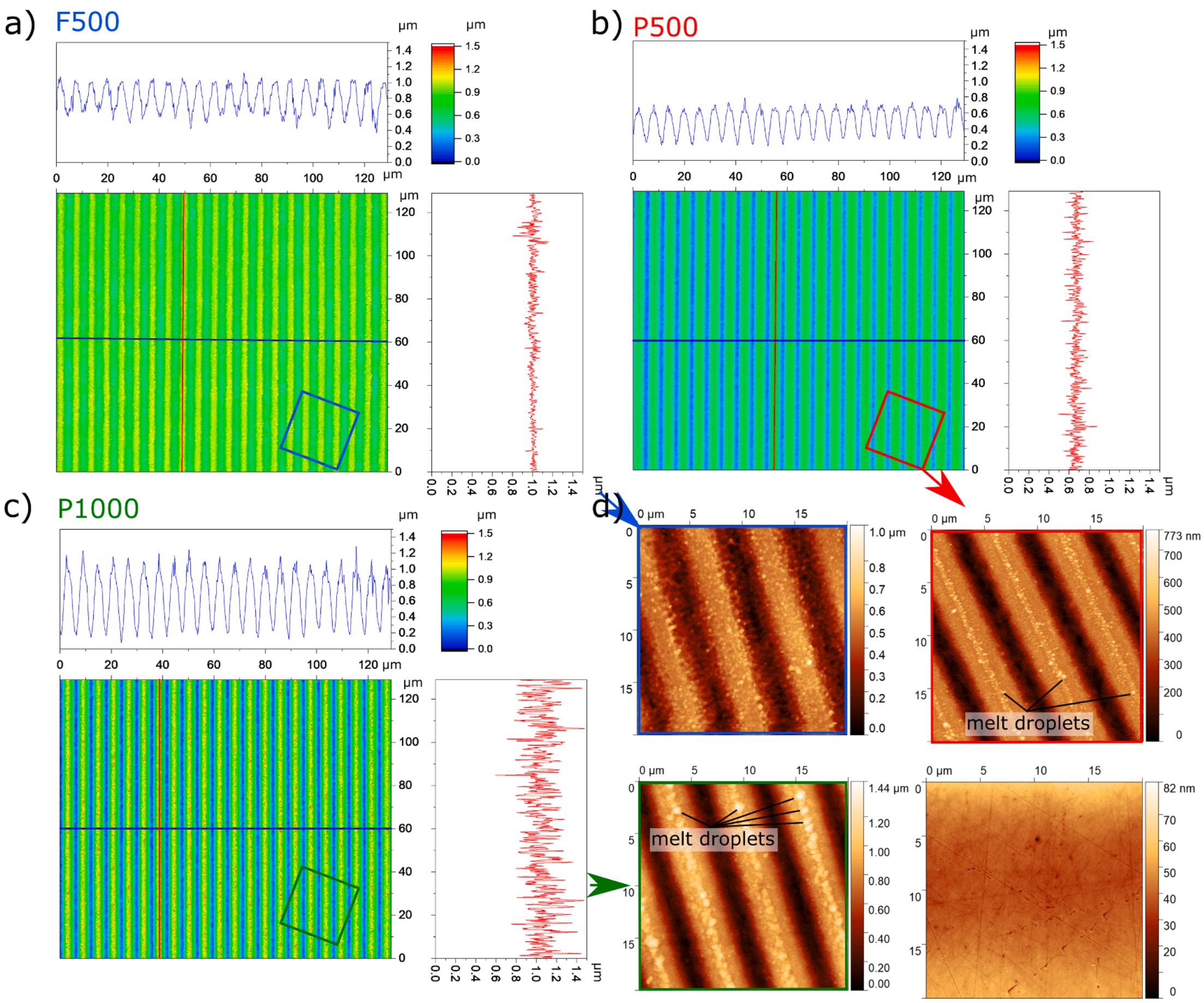

**Fig. 11.** Topographical analysis of the laser samples and the polished reference. (a – c) LSM-Scans of the DLIP samples with profile plots parallel and perpendicular to the main structure (F500 = fs-DLIP, 500 nm deep / P500 = ps-DLIP, 500 nm deep / P1000 = ps-DLIP, 1000 nm deep). (d) AFM-scans of the three DLIP-structures and the polished reference sample.

**Table 3**
Results of roughness analysis by LSM ($R_c$, $S_a$ and S-ratio) as well as AFM (S-ratio AFM) for the four sample types.

| | F500 | P500 | P1000 | Reference |
|---|---|---|---|---|
| $R_c$ [nm] | 474 ± 22 | 471 ± 14 | 1005 ± 27 | – |
| $S_a$ [nm] | 147 ± 3 | 134 ± 3 | 272 ± 1 | 4 ± 0 |
| S-ratio | 1.068 ± 0.002 | 1.038 ± 0.001 | 1.142 ± 0.002 | 1 ± 0 |
| S-ratio AFM | 1.132 | 1.122 | 1.192 | 1.001 |

The characterization of the F500 samples in Fig. 12 a) – d) reveals topographic maxima with only a discreet secondary roughness dominated by flakelike structures reported in another study as well on DLIP-copper samples with a periodicity of 3 µm [45]. The valleys of the structures depict a pronounced roughness as a network of crater structures forms a sub-pattern (compare Table 4). The P500 surface differs mainly from the F500 surfaces through the agglomeration of oxidic particles and melt droplets arranged in a line along the structural maxima increasing local roughness and decreased roughness in the valleys (Rq = 30 nm ± 4). The FIB cross-section in Fig. 12 g) reveals a similar profile line as the F500 samples except of the decoration with oxidic features on the structure peaks. For the deeper ps samples P1000, the agglomeration of oxidic particles and especially melt structures is more pronounced than on the P500 samples. Oxidic nanoparticles are not flakelike as observed on the F500 samples but show a more regular morphology. The FIB cross-section reveals a more sinusoidal profile line due to the stronger agglomeration of oxidic particles and melt structures on the peaks of the pattern. An increase in pulse duration from the fs to the ps-regime leads to a higher temperature impact on the copper surface. It has been reported that resolidified particles appear for ps laser treated copper surfaces along oxide layering while fs treated copper only showed a thin oxidic layer on top of the copper substrate [60]. The larger heat effected zone of the substrate treated with a ps laser pulse was accompanied by a more pronounced recast area as well as ejected and resolidified particles on aluminum [61,62], which is in good agreement with our observations for DLIP on copper. Regarding the sub-pattern morphology the more defined crater structures could be explained by the findings by Winter et al. [63]. They observed that copper irradiated with 525 fs builds a stronger "eruption structure" in the sub-µm range while the 20 ps irradiated copper showed increased bubble sizes in the µm range. Though this is expected to increase the roughness for the ps samples, the F500 samples show a finer network forming in the valleys, which might be overlayed by more pronounced melt structures for the P500 sample. Rajan et al. [61] reported stronger recast deposition after laser ablation with higher ps powers on aluminum and we can confirm these results for DLIP on copper (compare P1000).

XRD measurements were conducted in the range where oxidic copper peaks (CuO and $Cu_2O$) are expected and are summarized in Figure S2. Though a clear peak assignment to reference spectra was not possible, a general trend could be observed supporting the SEM findings with more oxides for F500, P500, and P1000, respectively. For near-surface characterization of the chemical composition of the surfaces, larger scale XPS analysis, that does not resolve the laser structures was conducted. Fig. 13 shows the elemental concentrations of the surface

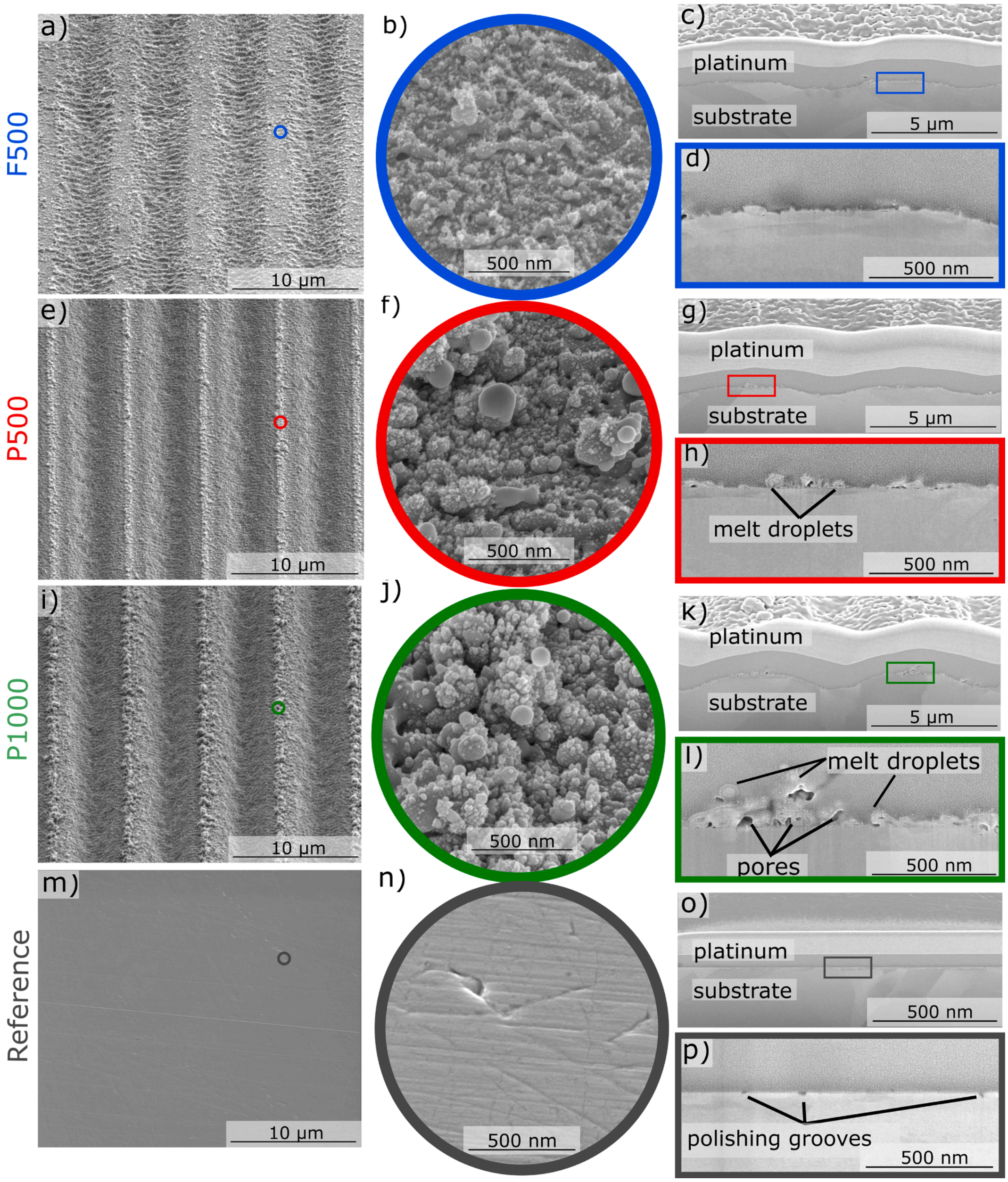

**Fig. 12.** SEM- and FIB characterization of the four sample types (F500 = fs-DLIP, 500 nm deep / P500 = ps-DLIP, 500 nm deep / P1000 = ps-DLIP, 1000 nm deep): (a) – (b) SE images of F500 with mild nanoroughness on the peaks and pronounced roughness in the structure valleys. (c) – (d) FIB-cross section of F500 showing the flattened peaks. (e) – (f) SE images of P500 with more pronounced nanoroughness on the peaks through melt droplets. (g) – (h) FIB-cross section of P500 showing the flattened peaks decorated with melt droplets and nanoparticles. (i) – (j) SE images of P1000 with strong nanoroughness on the peaks through melt droplets and oxidic particles. (k) – (l) FIB-cross section of P1000 showing a more sinus-like profile and a strong decoration of the structure peaks with melt droplets enclosing pores with the substrate. (m) – (n) SE images of the polished reference with some artefacts from the mechanical polishing. (o) – (p) FIB-cross section of the polished reference showing the flat profile with some polishing grooves.

**Table 4**
Roughness analysis of the peaks and valleys of the laser patterned samples.

| $R_q$ [nm] | Peak | Valley |
|---|---|---|
| F500 | 35 ± 5 | 51 ± 3 |
| P500 | 51 ± 11 | 30 ± 4 |
| P1000 | 123 ± 15 | 37 ± 3 |

composition of the DLIP-samples and a polished copper surface after aging, core level spectra can be found in Figure S3. The elemental concentrations shown in Fig. 13 a) rule out unexpected foreign contamination. The detected Cu binding states according to the results of the analysis of Cu LMM spectra shown in Fig. 13 b) indicate that $Cu_2O$ forms the oxidic layer for all sample types. The analysis according to the Cu2p spectra and the detail O1s spectrum confirm the domination of $Cu_2O$ over CuO. The detail O1s spectrum additionally shows a strong signal of organic bound oxygen due to the organic adsorption layers. Fig. 13 c) depicts the detected C binding states analyzed from the C1s spectrum. Due to sample aging, a strong adsorption of aliphatic carbon groups appears on all sample types with a minor contribution of polar groups. The amount of adsorbed carbonous groups measured as C/Cu ratio is given in Table 5. The ratios are smaller than expected from Fig. 2 for samples aged for 12 weeks but all samples are still in the regime where stabilized SCAs were observed on polished surfaces (compare

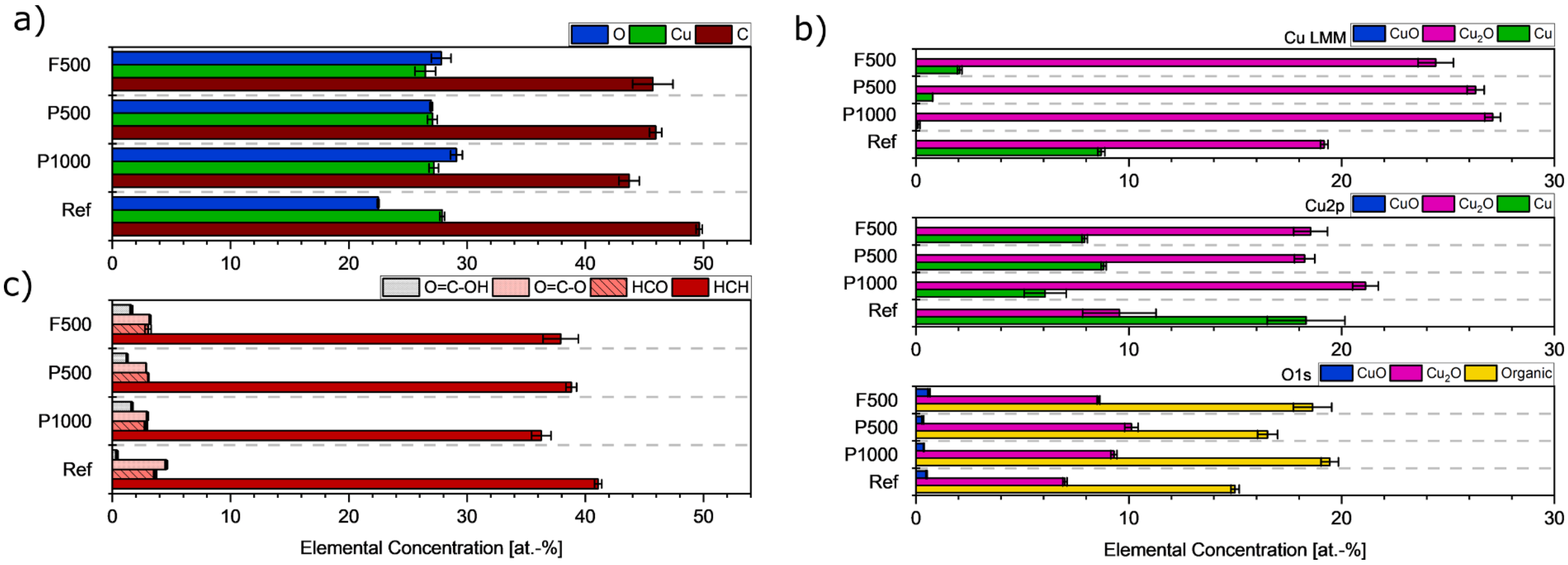

**Fig. 13.** XPS overview spectrum (a) and detail spectra Cu LMM, Cu2p and O1s (b) and C1s (c) of the DLIP-samples and a polished reference sample after aging. A strong hydrocarbon adsorption layer covers the copper substrate with its oxidic reaction layer dominated by $Cu_2O$.

**Table 5**
The C/Cu ratio serves here as a measure for the adsorption of volatile hydrocarbon groups and ratio $C_{pol}/C_{nonpol}$ describes the polarity ratio of the adsorption layer for all sample types.

| XPS ratios | F500 | P500 | P1000 | Reference |
|---|---|---|---|---|
| C/Cu | 1.73 ± 0.12 | 1.7 ± 0.04 | 1.61 ± 0.05 | 1.78 ± 0.02 |
| $C_{pol}/C_{nonpol}$ | 0.21 ± 0 | 0.18 ± 0 | 0.21 ± 0.01 | 0.21 ± 0.01 |

Fig. 2). Reasons could be another batch of the packaging paper that adsorbed less carbonous species itself but also fluctuations in humidity or temperature during sample preparation could potentially influence the adsorption process. $C_{pol}/C_{nonpol}$ delivers very comparable values for all sample types. Fig. 13 c) reveals the presence of COOH groups on the laser treated samples which are barely present on the polished reference (see Figure S3 for details). They were also found in a previous study of laser treated copper surfaces [18,45] and might be adsorbed during the laser process itself by interactions of the laser plasma with the surrounding atmosphere. The results of XPS analysis depicted in Fig. 13 and Table 5 reveal a dominance of $Cu_2O$ for all sample types regarding the oxide state and a strong hydrocarbon adsorption with comparable adsorption film compositions.

To increase information depth of the XPS analysis, sputter profiles were recorded for all sample types (compare Fig. 14) supporting previous results by showing a more pronounced oxide layer from F500 to P500 and P1000 and a comparable course of the C concentration.

To summarize the XPS analysis, the sample types differ in the amount of surface oxide, respectively oxide layer thickness, but not in the dominating oxide type. Also, all samples show pronounced uptake of adventitious carbon and the adsorbed hydrocarbon film covering the oxide shows a very comparable composition. Due to the coverage of all samples with the adsorption film, differences observed in the wetting behavior can be assigned to the variations in topography. Fig. 15 summarizes the wetting results of the SCA tests on the different surfaces with top-view images of the droplets. As can be seen from Fig. 15 a), the F500 samples show very high SCAs of approximately 170° parallel and perpendicular to the line pattern with no visible anisotropy. The wetting state seems stable over time and no spreading of the droplet was observed. Though the P500 samples are characterized by the same primary topography, a completely different wetting response to the applied water droplets can be observed. It is characterized by a strong wetting anisotropy with SCAs of approximately 160° perpendicular to the line pattern and 130° in the other direction. A slight settlement of the droplet along the roughness is indicated by the gentle decrease of the SCA after 60 s. The behavior of the P1000 samples is comparable to the F500 samples with SCAs in the range of 170° perpendicular to the line pattern. A slight anisotropy of approximately 5° can be observed that seems to mildly increase with time, but it must be mentioned that droplet deposition on the P1000 samples posed a challenge despite the increased volume of 6 μl and took more time which is why the slight decrease of SCAs after deposition could also be due to a modest droplet evaporation. As XPS analysis revealed that all sample types are covered by an adventitious carbon film that is in direct contact to the water in the wetting experiments, topography can be drawn responsible for the observed wetting differences. Especially the anisotropic wetting of the P500 samples compared to the F500 samples indicates that the same primary topography may induce different wetting regimes when combined with varying secondary roughness features. Based on previous findings [27] and ref. [64], we expect a CB wetting state for line patterns that do not demonstrate wetting anisotropy and a Wenzel wetting state for the P500 samples that do demonstrate wetting anisotropy. Thereby, wetting anisotropy is induced by the droplet entering the pattern, where capillary forces can pull the droplet along the pattern grooves. In order to better observe the existing wetting states, images of additionally applied droplets were taken in top-view with a digital microscope (compare Fig. 15 b)) without the cover set up.

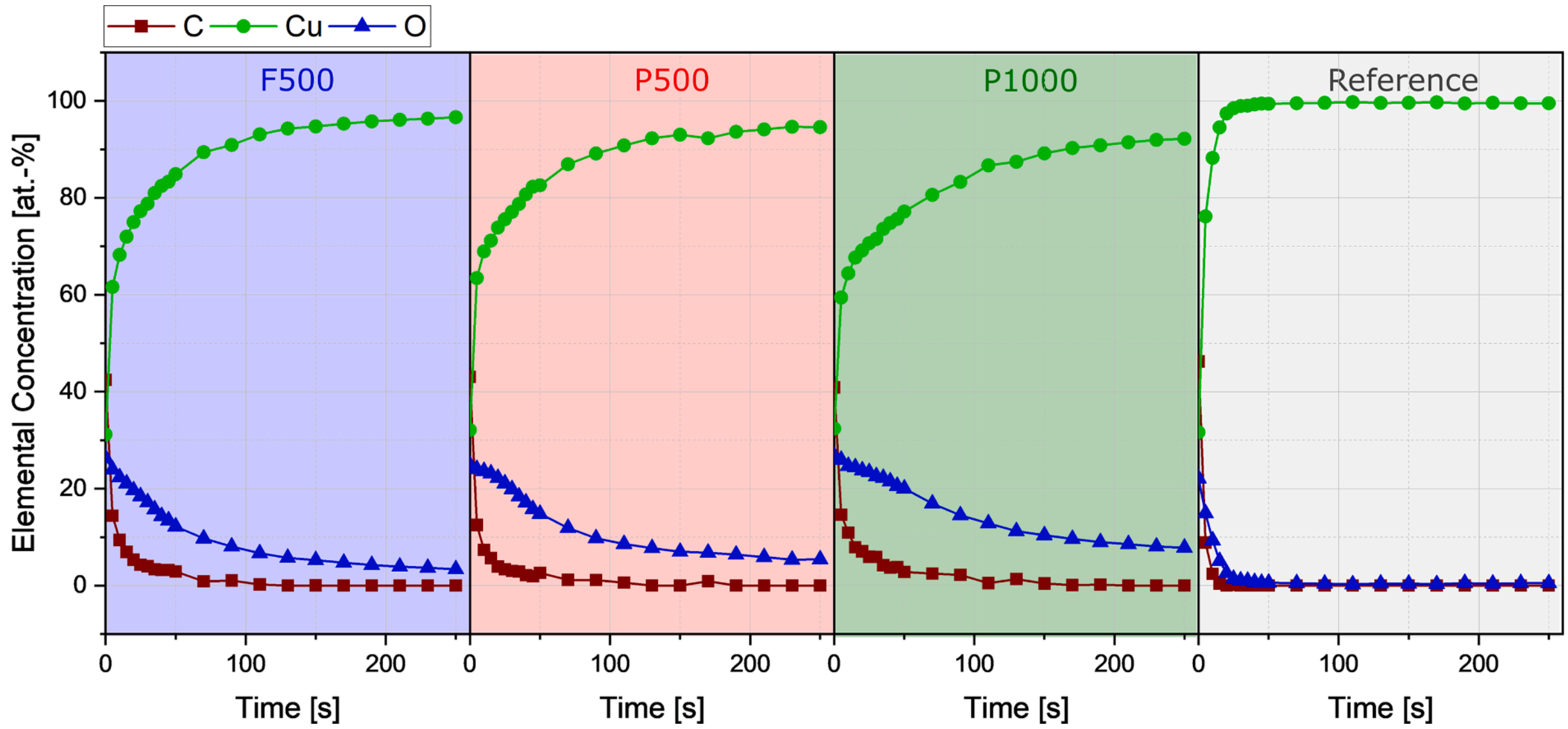

**Fig. 14.** In order to map the depth distribution of the detected elements, XPS sputtering depth profiles were recorded. The most pronounced difference can be found in the oxygen concentration falling abruptly on the polished reference on the one hand and decreasing progressively on the laser treated surfaces on the other hand. Results indicate a stronger oxide formation from F500 to P500 and P1000. The course of the C concentration is comparable for all sample types.

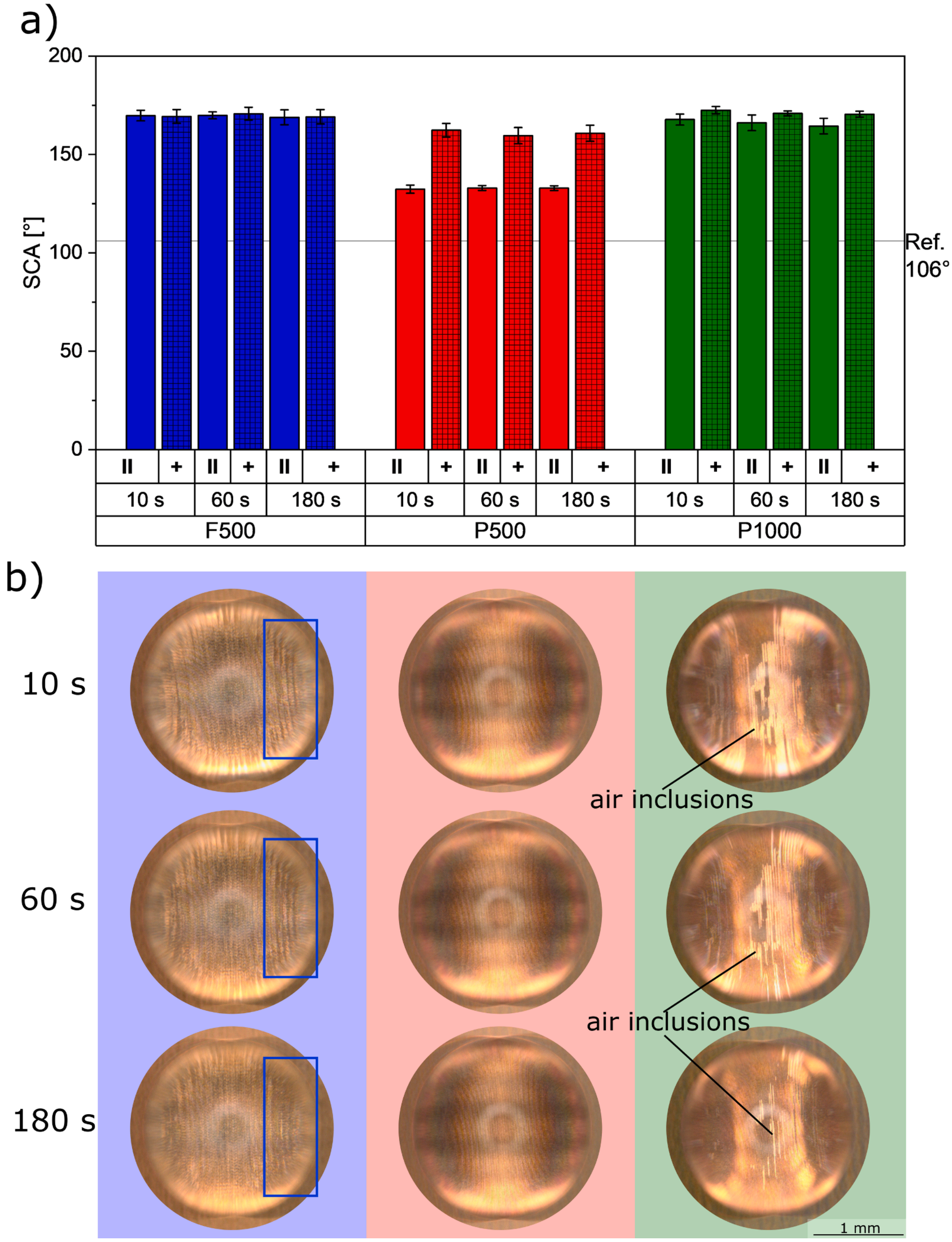

**Fig. 15.** (a) Mean value of SCAs of aged DLIP samples (F500 = fs-DLIP, 500 nm deep / P500 = ps-DLIP, 500 nm deep / P1000 = ps-DLIP, 1000 nm deep) 10 s, 60 s and 180 s after droplet deposition parallel (II) and perpendicular (+) to the line-like pattern with the standard deviation. Measurement conditions: ≈ 22 °C; ≈ 49 % RH.; at least 9 droplets on 3 samples for 10 s measurement per type; 6 droplets on 3 samples for 60 s/180 s measurement per type. (b) Top view of the droplets on the different surfaces. Bright stripes indicate air bubbles trapped in the valleys of the structure due to the change in reflectivity.

Images of droplets on the F500 sample show subtle visible brighter stripes in the line pattern which are assumed to be caused by trapped air bubbles. After 180 s, margins of the droplet show less bright stripes indicating a partial transition to the Wenzel state (compare blue frame). The P500 sample depicts a Wenzel wetting state already 10 s after droplet deposition without any visible air inclusions, which matches the observed wetting anisotropy. A strong spreading is not observed. The P1000 samples show air inclusions due to their increased aspect ratio compared to P500 [56,61,65]. Here, the transition to the Wenzel state is pronounced moving from the outside to the inside of the droplet, which is in good agreement with the findings by Murakami et al. [56]. Still, no strong deviation in SCA with time occurs possibly due to droplet pinning on the increased peak roughness. Wetting of the secondary peak roughness cannot be assessed but concerning the primary pattern, the F500 samples show the most stable CB state, followed by the P1000 samples with a clear transition to the Wenzel state within a few minutes and the P500 state with a stable Wenzel wetting already after 10 s. For pattern grooves smaller than the capillary size of water (2 mm), elongation of the droplet is expected in the Wenzel state due to capillary forces reducing with increasing surface hydrophobicity and CB wetting [64]. We barely observe droplet elongation in top view for P500 despite apparent Wenzel wetting which we trace back to the high chemical hydrophobicity and a complex interplay with surface topography, e.g. droplet pinning. Our results indicate that not only the primary topography (line pattern) and its aspect ratio, but also the sub-pattern morphology play a major role regarding the wetting state due to the strong difference between F500 and P500. We believe that the increased roughness in the valleys of F500 samples favors air inclusions and isotropic wetting. We want to highlight that the formation of periodic sub-patterns in the valleys like LIPSS instead of the here observed random network structures might trigger pronounced wetting as capillary forces could be active on two scales (primary and secondary pattern). Detailed investigations on the effect of periodic sub-pattern morphologies and their orientation relative to the primary pattern will be necessary in the future.

Analysis of water adhesion by measurements of advancing contact angle (ACA) and receding contact angle (RCA) was not found suitable for our samples (details can be found in the supplementary material). In order to still be able to observe whether the samples show a high or a low water adhesion, tilting experiments were conducted parallel and perpendicular to the line pattern. Due to the use of a prototype tilting stage, we refrain from giving exact sliding angles and interpret the results in order to assign the observed wetting behavior to a low water adhesion as observed for the lotus effect or a high adhesion as shown by the rose petal effect [65]. Images of the tilted surfaces are shown in Figure S4. F500 as well as P1000 samples showed rolling of the droplet independently of the tilting direction, while the droplet stuck on the P500 samples and the reference samples even after a tilt of 180°. On the F500 sample the water leaves occasional residues. On the P1000 sample only very small residues are left, which evaporate within seconds after the rolling of the droplet. High water adhesion as observed on P500 can be caused by penetration of the primary roughness with (Wenzel) or without penetration of the secondary peak roughness (Cassie impregnating wetting state) as postulated for the rose petal effect [66,67]. Rose petal behavior was already observed on DLIP patterns on steel [68] as well as on flat fs-treated copper transitioning to an increasing lotus like behavior for more pronounced structures [65] which resembles the trend we observe between P500 and P1000.

Bhushan et al. [67] depicted a wide range of possible wetting regimes for hierarchical microstructures with a primary roughness and a secondary nanoroughness. They distinguish between a lotus effect with air inclusions with the primary and secondary roughness and a Cassie state penetrating only the nanoroughness on top of the primary topography. For the F500 and the P1000 samples the water residues indicate adhesion or pinning to some topographic features whose characteristics cannot be determined in detail in our experimental setup. We expect a further adsorption of hydrocarbons (e.g. due to continued aging) or the deposition of more hydrophobic molecules to reduce the water residues on the surfaces of F500 and P1000 and to further lower their sliding angle as higher concentrations of nonpolar surface groups are expected to increase and further stabilize trapped air and delay penetration of the water into the roughness. This is also believed to fully cancel any anisotropy on P1000 but otherwise not expected to majorly influence the SCAs. For the P500 samples, it is uncertain at this point whether a more hydrophobic surface termination might reduce the witnessed anisotropy and favor air inclusions as witnessed for F500 or if the topography prevents CB wetting independent of chemical hydrophobicity. More studies in the future will be necessary to address this question. The application of hydrophilic coatings is expected to greatly enhance the wettability of all surfaces, transferring all patterns into the Wenzel wetting state with a strong water adhesion. However, detailed analysis of these research questions was beyond the scope of this work and will be required in the future to verify these assumptions.

Clearly, very slight changes in topography can induce significantly different wetting behavior. Though an exact determination of the Wenzel factor $r$ and the CB factor $f$ are not possible in our experiments, indications exist regarding the wetting states. We assume that the pronounced roughness in the valley of the primary topography of the F500 samples in combination with the flattened peaks impedes the droplet from wetting the valleys, induces air-inclusions and therewith results in the roll-off behavior. The subtle decoration of the peaks with flakelike nanoparticles might push the wetting behavior even further to the superhydrophobic state [39]. A slow transition towards Wenzel state was observed. The same primary topography produced by ps-DLIP (P500) with an increased roughness on the structural peaks due to a stronger agglomeration of melt structures and a decreased roughness in the valleys induces wetting of the valleys which also causes the witnessed anisotropy of the SCAs and a strong water adhesion. Sticking of the droplet is as pronounced as on the polished reference and supports the idea of either a full Wenzel or a Cassie impregnating wetting state as shown in ref. [66]. The higher structures of the P1000 samples induce air inclusions [56,65] despite increasing peak roughness, although no stable transition is observed. Still, the present air inclusions prevent the droplet from sticking to the surface when turned upside down. Clearly, topography has a major influence on the wetting regime and even with the same primary pattern, different wetting states can be achieved.

## 4. Conclusions

This study provides a comprehensive insight into the wettability of copper with its multi-step experimental approach. The precise design of the experiments and the comprehensive chemical as well as topographical analysis allow for the clear separation of both effects on the wetting behavior of copper. New insight is gained into the ageing behavior of copper and the formation of carbonous adsorption films and applied to interpret the impact of topography on different scales. The adsorption of organic species from the atmosphere is known to heavily influence the resulting wetting behavior [18]. In the first set of experiments, we were able to show that upon controlled aging of polished copper surfaces in contrast to ambient storage conditions the SCA stabilizes despite ongoing increase in hydrocarbon concentration and fluctuations in the relative amount of polar adsorbates. This supports our findings from a previous study [20] that a material- and microstructure dependent threshold of adsorbed hydrocarbons might exist after which additional adsorption only leads to gradual increase of SCA. This is in agreement with the simulative findings by Korczeniewski et al. [21] which showed that once a monolayer of hydrocarbon is formed, additional adsorption does not induce strong changes in SCA anymore, while we were able to not only determine the amount of adsorbed hydrocarbons but also the binding states and the arrangement in detail by XPS-analysis. Controlled adsorption experiments will be needed in the future to elucidate the role of adsorption layer density and thickness regarding the wetting response. Our results suggest that after chemisorption which we draw responsible for the wetting transition, physisorption might lead to a further growth of the adsorbate layer as no saturation of C/Cu was observed.

Based on these findings, samples in the following experimental sets were stored under strictly controlled conditions. In order to investigate the role of copper oxide in the wetting response of aged samples, magnetron sputter deposition was applied to form Cu, CuO and $Cu_2O$ on polished copper substrates. Though XPS revealed a more pronounced hydrocarbon adsorption for Cu, SCA analysis delivered similar long-term wetting behavior for all three surface types. This finding goes along with a study published recently [26] where rough copper oxides produced by hydrothermalization showed the same hydrophobicity after storage in C-rich atmospheres for different times. While in reference [26], the impact of the copper oxide state was analyzed on topographically altered surfaces, we performed the investigations on flat samples to exclude an overlaying impact of a pronounced roughness. We kept storage conditions the same and minimized the influence of topography and therewith can disprove the hypothesis that a deoxidation form CuO to $Cu_2O$ might be responsible for a wetting transition [31] as both oxides turn hydrophobic upon exposure to atmospheres rich in hydrocarbons. Based on these findings, the influence of topography on the wetting behavior of copper was investigated in detail by surface laser treatment. In order to understand the role of a mild isotropic roughness first, LR was implemented to uniformly roughen the surface. An increase of SCA of 25 % compared to the polished surface was reached. Again, identical storage conditions were applied to surface types under direct comparison to each other making our estimation of 25 % reliable. A strong hydrocarbon layer was detected, which allowed the topography to show its potential for further hydrophobization. For the analysis of hierarchical structures, two different pulse durations were applied in the DLIP set up to generate the same primary line-like topography with varying sub-pattern morphologies. Contrary to other studies taking different pulse durations into account [62], we aimed at keeping the primary topographies as similar as possible. As the oxide state proved to not influence the wetting in the aged state and samples were equally stored to form adsorption layers, clearly topography could be drawn responsible to the distinct difference in the wetting behavior. A more pronounced roughness in the structural valleys in combination with minor topographic features on the peaks in from of a mild decoration with flakelike nanoparticles favors air-inclusions and roll-off behavior. More pronounced roughness on the peaks induced by melt structures with reduced topography in the valleys rather induces wetting of the valleys and strong adhesion. Increasing the depth of these structures can induce air inclusions and cause rolling of the droplet despite increasing peak roughness. While a variation of the primary topography and especially the aspect ratio proved to influence water adhesion before [56,61,65], we were able to induce opposite wetting phenomena while maintaining the primary topography. A stronger hydrophobic surface termination potentially further stabilizes CB wetting which might affect especially the sliding behavior towards smaller angles. Detailed experimental studies of the dependence of the dynamic wetting behavior of hydrocarbon adsorption will be needed in the future. Also, different sub-pattern morphologies e.g. LIPSS need to be investigated in detail in systematic experimental approaches.

With these results, our work provides a comprehensive overview of the wettability of copper.

The knowledge gained in the presented study gives fundamental insight into the wetting behavior of surfaces, particularly patterned samples and allows for the detailed design of copper surfaces targeting the desired static as well as dynamic wetting behavior.

## CRediT authorship contribution statement

**Sarah Marie Lößlein:** Writing – review & editing, Writing – original draft, Visualization, Validation, Methodology, Investigation, Formal analysis, Conceptualization. **Rolf Merz:** Writing – review & editing, Writing – original draft, Visualization, Validation, Methodology, Investigation, Formal analysis. **Yerila Rodríguez-Martínez:** Writing – review & editing, Methodology, Investigation. **Florian Schäfer:** Writing – review & editing, Methodology, Investigation. **Philipp G. Grützmacher:** Writing – review & editing, Conceptualization. **David Horwat:** Writing – review & editing, Supervision, Resources. **Michael Kopnarski:** Supervision, Resources, Project administration, Funding acquisition. **Frank Mücklich:** Supervision, Resources, Project administration, Funding acquisition.

## Declaration of competing interest

The authors declare that they have no known competing financial interests or personal relationships that could have appeared to influence the work reported in this paper.

## Data availability

Data will be made available on request.

## Acknowledgements

We would like to thank the German Research Foundation (Deutsche Forschungsgemeinschaft, DFG) for the financial support of the project number 435334669 (MU 959/50-1). We acknowledge the funding by the German Research Foundation (Deutsche Forschungsgemeinschaft, DFG) for both the fs laser (INST 256/562-1 FUGG) and the picosecond DLIP machine (INST 256/470-1 FUGG). We want to thank Rouven Zimmer from Chair of Functional Materials in Saarbrücken (Germany) for the support with the laser patterning of the LR-samples. We gratefully thank Michael Moseler and the team form Fraunhofer Institute for Mechanics of Materials IWM in Freiburg (Germany) – especially Kerstin Falk, Leonhard Mayrhofer, Prasanth Ganta and Claas Bierwisch – for the discussion, theoretical investigation and simulations of the observed phenomena. A manuscript evolving around the simulative interpretation of our experimental findings is under preparation by them.

## Appendix A. Supplementary data

Supplementary data to this article can be found online at https://doi.org/10.1016/j.jcis.2024.04.212.